\documentclass[aps,pre,a4,floatfix,groupedaddress,showpacs,showkeys,onecolumn]{revtex4}  
\usepackage{graphicx,amsmath,amssymb,dsfont,bm}
\usepackage{color}
\usepackage{enumitem}
\usepackage{boldline}

\usepackage[normalem]{ulem}
\usepackage{xcolor}
\newcommand\redsout{\bgroup\markoverwith{\textcolor{red}{\rule[0.5ex]{2pt}{0.6pt}}}\ULon}
\usepackage{multirow}

\begin{document}

\title{Microscopic Understanding of Cross--Responses between
  Stocks:\\ a Two--Component Price Impact Model }

\author{Shanshan Wang} \email{shanshan.wang@uni-due.de}
\affiliation{Fakult\"at f\"ur Physik, Universit\"at Duisburg--Essen,
  Lotharstra\ss e 1, 47048 Duisburg, Germany} 
\author{Thomas Guhr} \email{thomas.guhr@uni-due.de}
\affiliation{Fakult\"at f\"ur Physik, Universit\"at Duisburg--Essen,
  Lotharstra\ss e 1, 47048 Duisburg, Germany}

\date{\today}

\begin{abstract}
 We construct a price impact model between stocks in a correlated market.  For the price change of a given stock induced by the short--run liquidity of this stock itself and of the information about other stocks, we introduce a self-- and a cross--impact function of the time lag. We model the average cross--response functions for individual stocks employing the impact functions of the time lag, the impact functions of traded volumes and the trade--sign correlators. To quantify the self-- and cross--impacts, we propose a construction to fix the parameters in the impact functions. These parameters are further corroborated by a diffusion function that measures the correlated motion of prices from different stocks. This construction is mainly ad hoc and alternative ones are not excluded. It turns out that both the sign cross-- and self--correlators are connected with the cross--responses. The self-- and cross--impact functions are indispensable to compensate amplification effects which are due to the sign correlators integrated over time. We further quantify and interpret the price impacts of time lag in terms of temporary and permanent components. To support our model, we also analyze empirical data, in particular the memory properties of the sign self-- and average cross--correlators. The relation between the average cross--responses and the traded volumes which are smaller than their average is of power--law form.

\end{abstract}


\maketitle

\section{Introduction}
\label{section1}

The price impact, \textit{i.e.} price response to trades, quantifies the expected price change conditioned on initiating a trade with a given size and a given sign (buy or sell)~\cite{Bouchaud2004,Moro2009,Farmer2013}. A buyer--initiated trade is expected to push the price up, while a seller--initiated trade to drop the price down, as from an economic perspective an increase in the demand should raise the price, and an increase in supply should reduce the price. In the view of market microstructure, the price change corresponding to the buyer-- or seller--initiated trades is due to the market orders annihilating all the limit orders in the best quote so that to shift the price to a higher buy price or a lower sell price. Since the price impact has been studied numerously, it is not difficult to find out its importance either in practice or in theory. In practice, to profit from trades as much as possible, it is essential to estimate the trading costs~\cite{Gatheral2010}, which are partly due to the price impact regarding to the brokerage commission. Thus, against the high costs induced by the price impact, the extremely large market order is restricted. In theory, by minimizing such costs, one can obtain the optimal execution strategies with the principle of no dynamic arbitrage or no price manipulation~\cite{Gatheral2010,Gatheral2012,Gatheral2013,Obizhaeva2013,Alfonsi2014,Alfonsi2016}, where the functional form of price impact is necessary to be given. As the price change relates to the demand and supply of the stock~\cite{Hopman2007}, the impact reflects the shape of excess demand in economics.

The price impact is mainly generated by three factors out of all possible ones, these are the short--run liquidity costs, the traded volumes and the effects due to information~\cite{Chan1993,Farmer2004,Weber2006,Joulin2008,Moro2009,Hopman2007,Farmer2007}. In computerized financial market, the market liquidity is self--organized, that is to say, any trader can choose to be liquidity taker or liquidity provider. Not only the market markers in the traditional sense provide liquidity. As liquidity providers, the traders post limit orders to sell or buy certain volumes at a minimum ask price or a maximum bid price. To execute a transaction immediately, the traders, as liquidity takers, consume liquidity by launching market orders to buy or sell certain volumes. The costs for making transactions without time delay are referred to as liquidity costs~\cite{Demsetz1968,Bartram2008}. As a part of the trading costs, they are to be distinguished from the fixed brokerage commission. They are to some extent characterized by the bid--ask spread~\cite{Demsetz1968,Bartram2008,Wyart2008} which measures the cost of an instantaneous round--trip of one share, \textit{i.e.} a buy instantaneously followed by a sell. The short--run liquidity costs arise from the difficulty to immediately find buyers or sellers. Hence, price concessions occur to increase the volumes being traded instead of attracting more buyers or sellers~\cite{Chan1993}. Since the market orders with large volumes move the price considerably, the impact of traded volumes has drawn much attention. Although various relations between the price changes and the traded volumes were found~\cite{Potters2003,Lillo2003}, most studies~\cite{Farmer2004,Weber2006,Joulin2008,Moro2009,Bouchaud2004} demonstrate that large price changes are, on average, not the result of large traded volumes. Rather, they are driven by the low density of limit orders stored in the order book, \textit{i.e.} by the small liquidity. The less the liquidity, the larger is the gap of limit orders in the order book. It has been demonstrated in an agent based model~\cite{Schmitt2012} that such gaps lead to large price shifts. The resulting heavy tails of the return distributions are one of the celebrated stylized facts and specific features in price dynamics~\cite{Cont2001,Chordia2002,Bouchaud2003,Bouchaud2009,Chakraborti2011,Toth2011,Eisler2012,Schmitt2012,Schmitt2013}. Large market orders are often thought by the other traders to carry some specific information. No matter, whether they are really driven by valid information, the large orders alert the other traders and make them adjust their strategies. Thus, to conceal the trading information and minimize their trading costs, the traders who try to issue large orders prefer to split them into small pieces~\cite{Bouchaud2004,Bouchaud2009,Toth2015}, implying a long--memory correlation of trade signs~\cite{Lillo2004,Bouchaud2004}. The orders which are small due to the splitting are thereby responsible for the large price change on average~\cite{Bouchaud2004}. The various effects of information always attract the academic attention~\cite{Hasbrouck1988,Grossman1976,Farmer2007,Joulin2008,Gomes2015}. Information can be classified as all the available public information, such as news, and the information which is often called private. The Efficient Market Hypothesis (EMH)~\cite{Fama1970} states that all available information is processed and encoded in the current prices, which would rule out any (statistical) arbitrage opportunity. Thus, if we leave aside, \textit{e.g.}, drastic political events, all the available public information in total does typically not generate large price change. This is consistent with the finding that neither idiosyncratic news nor market--wide news can explain the frequency and amplitude of price jumps~\cite{Joulin2008}. The price impact is more likely to be prompted by private information that is hidden in the trades and is then permanently incorporated in the new equilibrium price~\cite{Chan1993}.

An early model of the price impact was put forward by Kyle~\cite{Kyle1985}. It assumes a linear dependence of the impact on the traded volumes. However, subsequent studies find the impact is nonlinear in the traded volumes, described either by a power--law~\cite{Lillo2003,Farmer2004b,Almgren2005} or by a square--root~\cite{Torre1997,Gabaix2003,Plerou2004} or by a logarithm~\cite{Potters2003}. The price impact is thus seen to be permanent and fluctuates in order size~\cite{Lillo2004,Gerig2008}. Nevertheless, another opinion proposed by Bouchaud \textit{et al.}~\cite{Bouchaud2004} states that the price impact is neither linear in the volume nor permanent in time. It should rather be transient and fixed with order size. To be specific, these authors constructed a price impact model depending on time, where a `bare' impact function is used to propagate the impact of a single trade. The bare impact function is due to the strong self--correlations of trade signs and, in turn, offsets the amplification effect due to the time--accumulated trade sign correlation by decaying with time. The transient property of price impact is thus seen as a necessary consequence of the long--memory nature of the order flow~\cite{Bouchaud2004,Bouchaud2010}. Actually, the two opinions are not contradictory, neither in the functional forms~\cite{Gerig2008} nor in the time-decay mechanisms~\cite{Farmer2006}.

Importantly, the price impact models mentioned above are all confined to single stocks. That means the expected price change is only interpreted as a consequence of the trades in the same stock rather than influenced by the trading of other stocks. Recently, the price impact of trades, \textit{i.e.} the price cross--responses to trades, between different stocks in correlated financial market was studied empirically for individual stock pairs~\cite{Wang2016a} and also for suitable averages over stock pairs~\cite{Wang2016b}. As none of the available models accounts for the cross-responses, we here aim at extending the model of Bouchaud \textit{et al.}~\cite{Bouchaud2004}. While the latter contains one single bare impact function, we propose a price impact model for correlated markets which contains two impact functions depending on the time lag, one from the stock itself and the other one across stocks. As the terms suggest, the self--impact function propagates the impact of a single trade in the stock itself, while the cross--impact function transmits the impact of a single trade from another stock. The two impact functions result from the short--run liquidity of the stock itself and the trading information of other stocks, respectively. We demonstrate that the cross--response function of the price changes is indeed related to both, the self-- and the cross--correlations of trade signs.

The paper is organized as follows. In Sec. \ref{section2}, we construct the price impact model with two impact functions depending on the time lag, \textit{i.e.}, a self--impact function and a cross--impact function. We thereby obtain the average cross--response functions of individual stocks for three scenarios: the cross responses related to sign cross--correlations, to sign self--correlations, and to both, respectively. In Sec. \ref{section3}, we resort to empirical data to analyze and calibrate the memory properties of the trade sign self-- and cross--correlators, and to determine the relation between the average cross--responses and those traded volumes that are smaller than their average. To quantify the price impacts that are difficult to obtain from empirical data, we propose a construction to fix the parameters in the impact functions in Sec. \ref{section4}. In Sec. \ref{section5}, we introduce a diffusion function that describes the correlated motion of different stock prices, and use it to corroborate the parameters. In Sec. \ref{section6}, we quantify and interpret the price impacts of individual stocks in detail. We give our conclusions in Sec. \ref{section7}. Upon completion of this study, we became aware of related investigations~\cite{Benzaquen2016}.

\section{Price impact model}  
\label{section2}

We setup our model for the response functions in a market of correlated stocks.  In Sec. \ref{section2.1}, we collect the basic definitions as introduced in Refs.~\cite{Wang2016a,Wang2016b}, and in Sec. \ref{section2.2}, we construct the price impact model between two stocks to model the cross--response functions for stock pairs. In Sec. \ref{section2.3}, we reduce the complexity of the response functions to be averaged by defining them per share and also by restricting ourselves to three scenarios. We obtain the passive and active averaged response functions for the three scenarios in Secs. \ref{section2.4}, \ref{section2.5} and \ref{section2.6}, respectively.

\subsection{Basic definitions} 
\label{section2.1}

In the order book, a buy (sell) market order can either raise (lower) or leave unchanged the price at the best ask $a_i(t)$ (best bid $b_i(t)$). Both types of market orders move or leave unchanged the midpoint price
\begin{equation}
m_i(t)=\frac{1}{2}\left(a_i(t)+b_i(t)\right).
\label{eq2.1.1}
\end{equation}
Hence, the relative price change from time $t$ to time $t+\tau$ for the stock with index $i$, \textit{i.e.}, the logarithmic price difference or log--return is usually defined via the midpoint price,
\begin{equation}
r_i(t,\tau) =\log m_i(t+\tau) -\log m_i(t) =\log\frac{m_i(t+\tau)}{m_i(t)}.
\label{eq2.1.2}
\end{equation}
To distinguish the types of market order issued at time $t$, the trade sign 
\begin{eqnarray}
\varepsilon_j(t)=\left\{
\begin{array}{rl}
+1\ ,& \quad \textrm{for a buy market order}\ ,\\
 0 \ ,& \quad \textrm{for a lack of trading} \ ,	\\
-1\ ,& \quad \textrm{for a sell market order} \ , \\
\end{array}
\right.
\label{eq2.1.4}
\end{eqnarray}
is introduced.

The cross--response function for a stock pair $(i,j)$ measures the time dependent price change of stock $i$ triggered by either a buy or a sell market order of stock $j$ at time $t$. To acquire statistical significance, it is defined as the average
\begin{equation}
R_{ij}(\tau)=\Big\langle r_i(t,\tau)\varepsilon_j(t)\Big\rangle _t
\label{eq2.1.3}
\end{equation}
of the product of time-lagged returns $r_i(t,\tau)$ and trade signs $\varepsilon_j(t)$ over the time $t$. By definition, we have $i\neq j$ for the cross--response functions. We also introduce the time dependent trade sign cross--correlator between different stocks $i$ and $j$,
\begin{equation}
\Theta_{ij}(\tau)=\Big\langle \varepsilon_i(t+\tau)\varepsilon_j(t)\Big\rangle _t \ .
\label{eq2.1.5}
\end{equation}
We notice the properties
\begin{equation}
\Theta_{ij}(0)=\Theta_{ji}(0)
\qquad  \textrm{and} \qquad  
\Theta_{ij}(\tau)=\Theta_{ji}(-\tau) \ .
\label{eq2.1.6}
\end{equation}
Only for $i=j$, Eq.~(\ref{eq2.1.5}) becomes the auto--correlation of trade signs. 

In a correlated financial market, the price change of one stock $i$ might be impacted by many others simultaneously. Likewise, the trades of stock $i$ can also affect prices of several or many other stocks concurrently. By performing different averages over the stocks, we can shed light on both aspects.  We define a passive and an active average cross-responses,
\begin{equation}
R_i^{(p)}(\tau)\ = \ \left\langle R_{ij}(\tau)\right\rangle_j		
\qquad  \textrm{and} \qquad  
R_j^{(a)}(\tau)\ = \ \left\langle R_{ij}(\tau)\right\rangle_i \ ,
\label{eq2.1.7}
\end{equation}
where self--responses are excluded in the averages. The passive cross--response $R_i^{(p)}(\tau)$ measures how much, on average, the price of stock $i$ changes due to the trades of other stocks $j$, while the active cross--response $R_i^{(a)}(\tau)$ quantifies how the trading of stock $i$ influences the prices of other stocks $j$ on average. As shown in Ref.~\cite{Wang2016b}, the averages of cross--responses over different stock pairs reduce the response noise at large time lags . Correspondingly, we introduce the average passive and active trade sign cross-correlators
\begin{equation}
\Theta_i^{(p)}(\tau) \ = \ \left\langle \Theta_{ij}(\tau)\right\rangle_{j}
\qquad \textrm{and} \qquad
\Theta_i^{(a)}(\tau) \ = \ \left\langle \Theta_{ji}(\tau)\right\rangle_{j} \ ,
\label{eq2.1.8}
\end{equation}
excluding self--correlators in the averages.  They turn the short memory correlations for stock pairs into long memory effects, because large fluctuations are washed out~\cite{Wang2016b}.  The properties
\begin{equation}
\Theta_i^{(p)}(0) \ = \ \Theta_i^{(a)}(0)
\qquad  \textrm{and} \qquad 
\Theta_i^{(p)}(\tau) \ = \ \Theta_i^{(a)}(-\tau) \ 
\label{eq2.1.9}
\end{equation}
are worth to be mentioned.

\subsection{Setup of the model}
\label{section2.2}

Suppose a buy market order having a trade sign $\varepsilon_i(t)=+1$ with a unsigned volume $v_i(t)$ larger than that at the best ask was executed at initial time $t$. It is impossible to immediately issue new sell limit orders at the best ask in order to consume the volumes of the market order. In other words, there is an insufficient short--run liquidity. The buy market order therefore moves the initial trade price to a higher price instantaneously. Here, both the traded volume and the trade price are for the same stock $i$. The change of the trade price is reflected in the movement of the midpoint price, and the direction of the movement is indicated by the trade sign, \textit{e.g.} in the present case $\varepsilon_i(t)=+1$ for the price raising. In this study we use the logarithmic midpoint price to replace the trade price. The price impact from the traded volume is always non--negative either for a buy or sell market order of the same stock. It is denoted by $f_i(v_i(t))$. However, the price impact cannot persist all the time, as the new incoming limit orders enlarge the liquidity. We say that a stock is liquid if there are many shares which can be sold or bought without time delay and with little impact on the stock price. The liquidity can be estimated by looking at the bid--ask spread, \textit{i.e.} at the difference of the best bid and the best ask prices~\cite{Demsetz1968,Bartram2008}. Hence, if there is enough volume available at the new best ask with a price smaller than the one of last trade, the price in the following trades reverses. In view of the influence of the short-run liquidity, a price impact function $G_{ii}(\tau)$ versus time lag $\tau$ for a single trade is used to modulate the degree of the price impact due to the volumes traded. Furthermore, all other sources that indirectly cause the price change, such as the new information, are described by a random variable $\eta_{ii}(t)$. Hence, using discrete time, we have the trade price after the time step of length $\tau=1$,
\begin{equation}
\log m_i(t+1)=\log m_i(t)+G_{ii}(1)f_i\big(v_i(t)\big)\varepsilon_i(t)+\eta_{ii}(t).
\label{eq2.2.1}
\end{equation}
For the next time step, the trade price is not only influenced by the trade at the first time step, but also affected by the trade at the initial time $t$ with the remnant impact modulated by $G_{ii}(2)$ with $\tau=2$,
\begin{eqnarray} \nonumber 
\log  m_i(t+2)&=&G_{ii}(1)f_i\big(v_i(t+1)\big)\varepsilon_i(t+1)+\eta_{ii}(t+1) \\
  \nonumber
  &+&G_{ii}(2)f_i\big(v_i(t)\big)\varepsilon_i(t)+\eta_{ii}(t) \\
  &+&\log m_i(t) \ .
\label{eq2.2.2}
\end{eqnarray}
Now suppose infinitely many trades were executed before time $t$, each of these trades has an impact on the trade price at time $t$. Accounting for the past price at time $-\infty$, we obtain the trade price at time $t$ by constructing a superposition model, where all the price impacts from past trades are summed up,
\begin{eqnarray}  \nonumber
\log m_i(t)&=&\sum_{t'<t}G_{ii}(t-t')f_i\big(v_i(t')\big)\varepsilon_i(t')+\sum_{t'<t}\eta_{ii}(t') \\    
		&+&\log m_i(-\infty) \ .
\label{eq2.2.3}
\end{eqnarray}
We notice the sum over the random variables $\eta_{ii}(t')$. The prototype of this model was proposed in Ref.~~\cite{Bouchaud2004}. It describes the price impact from past trades, focusing on the same stock only. Here, we go beyond this and by comprising the trades from the same stock as well as from the other stocks.

When considering a trade from another stock $j$ with trade sign $\varepsilon_j(t')$, the trade also produces a price impact $g_i(v_j(t'))$ different from the one in the impacted stock $i$. As the trades of stock $j$ do not consume volumes directly from the order book of stock $i$, we attribute the price impact $g_i(v_j(t'))$ to transmission of trading information. We emphasize that the trading information in our model only contains trade directions, \textit{i.e.} buy and sell, and traded volumes of market orders, rather than other information, such as private information and relevant news which will later on be modelled by random variables. Due to the latter competing information, the price impact from volume traded for stock $j$ cannot remain unchanged. Hence, to scale how the price impact depends on the time lag, we employ a price impact function $G_{ij}(\tau)$ for a single trade. To distinguish these two types of impact functions, we refer to $G_{ii}(\tau)$ as to the self--impact function of the stock $i$, while we refer to $G_{ij}(\tau)$ as to the cross--impact function between impacted stock $i$ and impacting stock $j$. Moreover, we use random variables $\eta_{ij}(t')$ to model all the above mentioned sources belonging to stock $j$ that may cause price change of stock $i$. All the random variables $\eta_{ii}(t')$ and $\eta_{ij}(t')$ are assumed to be independent of trade signs and to not show autocorrelations in time. Thus, we arrive at the following model 
\begin{eqnarray}  \nonumber
\log m_i(t)&=&\sum_{t'<t}\Big[G_{ii}(t-t')f_i\big(v_i(t')\big)\varepsilon_i(t')+\eta_{ii}(t')\Big]  \\    \nonumber
		&+&\sum_{t'<t}\Big[G_{ij}(t-t')g_i\big(v_j(t')\big)\varepsilon_j(t')+\eta_{ij}(t')\Big]  \\
		&+&\log m_i(-\infty)
\label{eq2.2.4}
\end{eqnarray}
for the impacts of trades from different stocks.

As a consequence of the trade superposition model~\eqref{eq2.2.4}, the price change of stock $i$ resulting from Eq.~\eqref{eq2.1.2} comprises two components. The first one is due to the short--run liquidity of stock $i$ itself,
\begin{eqnarray}\nonumber
r_{ii}^{(L)}(t,\tau)&=&\sum_{t\leq t'<t+\tau}G_{ii}(t+\tau-t' )f_i\big(v_i(t')\big)\varepsilon_i(t')					\\  \nonumber
				   &+& \quad \sum_{t'<t}\Big[G_{ii}(t+\tau-t')-G_{ii}(t-t')\Big]f_i\big(v_i(t')\big)\varepsilon_i(t')	 \\  
				   &+&\sum_{t\leq t'<t+\tau}\eta_{ii}(t') \ .
\label{eq2.2.5}
\end{eqnarray}
As explained at the beginning of Sec.~\ref{section2.2}, the influence of the short--run liquidity is described by the self--impact $G_{ii}(\tau)$. Apart from the impact of traded volume, $G_{ii}(\tau)$ can be regarded as the impact of a single trade of stock $i$ on its own price after the time $\tau$. The second contribution results from the trading information transmitted from stock $j$ to stock $i$,
\begin{eqnarray}\nonumber
r_{ij}^{(I)}(t,\tau)&=&\sum_{t\leq t'<t+\tau}G_{ij}(t+\tau-t' )g_i\big(v_j(t')\big)\varepsilon_j(t') 				\\  \nonumber
				  &+&\quad \sum_{t'<t}\Big[G_{ij}(t+\tau-t')-G_{ij}(t-t')\Big]g_i\big(v_j(t')\big)\varepsilon_j(t')	 \\
				  &+&\sum_{t\leq t'<t+\tau}\eta_{ij}(t') \ .
\label{eq2.2.6}
\end{eqnarray}
Here, the cross--impact function $G_{ij}(\tau)$ plays the r\^ole of information propagator between stocks $i$ and $j$. It describes the impact of a single trade of stock $j$ on the price of stock $i$ after time $\tau$, without taking the impact of traded volume of stock $j$ into account. The sum of these two components
\begin{equation}
r_{ij}(t,\tau) = r_{ii}^{(L)}(t,\tau)+r_{ij}^{(I)}(t,\tau)
\label{eq2.2.7}
\end{equation}
constitutes the total price change of stock $i$ due to effects of stock $i$ and of another stock $j$.

Employing the definition \eqref{eq2.1.3}, we now calculate the time average response functions. For the two components of the price change of stock $i$ we obtain
\begin{eqnarray} \nonumber
R_{ij}^{(C)}(\tau)&=&\quad\Big\langle r_{ii}^{(L)}(t,\tau)\varepsilon_j(t)\Big\rangle _t		\\ 	\nonumber
		&=&\sum_{t\leq t'<t+\tau} G_{ii}(t+\tau-t')\left\langle f_i\big(v_i(t')\big)\right\rangle_t \Theta_{ij}(t'-t)\\ 
		 &+&\quad \sum_{t'<t}\Big[ G_{ii}(t+\tau-t')-G_{ii}(t-t')\Big]\left\langle f_i\big(v_i(t')\big)\right\rangle_t \Theta_{ji}(t-t')
\label{eq2.2.8}
\end{eqnarray}
\begin{eqnarray} \nonumber
R_{ij}^{(S)}(\tau)&=&\quad\Big\langle r_{ij}^{(I)}(t,\tau)\varepsilon_j(t)\Big\rangle _t		\\ 	\nonumber
		&=&\sum_{t\leq t'<t+\tau} G_{ij}(t+\tau-t')\left\langle g_i\big(v_j(t')\big)\right\rangle_t \Theta_{jj}(t'-t)\\ 
		&+&\quad \sum_{t'<t}\Big[ G_{ij}(t+\tau-t')-G_{ij}(t-t')\Big]\left\langle g_i\big(v_j(t')\big)\right\rangle_t \Theta_{jj}(t-t') \ .
\label{eq2.2.9}
\end{eqnarray}
The superscripts $(C)$ and $(S)$ refer to the response functions due to cross-- and self--correlations of trade signs, respectively. Thus, the price response of stock $i$ to the trades of stock $j$ contains both contributions,
\begin{equation}
R_{ij}(\tau) = R_{ij}^{(C)}(\tau) + R_{ij}^{(S)}(\tau) \ .
\label{eq2.2.10}
\end{equation}
Here we assume that both, the self-- and the cross--correlator of trade signs, are independent of the impacts of traded volumes, \textit{i.e.} of $f_i\big(v_i(t')\big)$ and $g_i\big(v_j(t')\big)$. The average impacts of traded volumes over all times $t$ do not depend on the traded time $t'$. We notice that Eq.~\eqref{eq2.2.8} only contains the trade sign cross--correlators $\Theta_{ij}(\tau)$ and $\Theta_{ji}(\tau)$, while Eq.~\eqref{eq2.2.9} only involves the trade sign self--correlators $\Theta_{jj}(\tau)$.

\subsection{Simplifications of the model}
\label{section2.3}

We want to perform two kinds of averages of the response functions with respect to the stock indices. To this end, we have a more detailed look at the functions \eqref{eq2.2.8} and \eqref{eq2.2.9} for a stock pair $(i,j)$ and we notice that the contribution of the traded volumes to the price response is independent of the time lag. This time independence means that the time--dependent information of the response is only included in the impact functions and the trade sign correlators. Since we focus on how the price changes respond to the trades on a certain time scale, it is reasonable to reduce the complexity of each component by dividing the average impact of traded volumes, after averaging the response components for an individual stock $i$ over different $j$, \textit{i.e.}, over the second index,
\begin{equation}
R_{i,0}^{(p, C)}(\tau) \ =  \ \frac{\left\langle R_{ij}^{(C)}(\tau)\right\rangle_{j}}{\Big\langle f_i(v_i)\Big\rangle_{t}}
\qquad \textrm{and} \qquad
R_{i,0}^{(p, S)}(\tau) \ =  \ \frac{\left\langle R_{ij}^{(S)}(\tau)\right\rangle_{j}}{\Big\langle g_i(v_j)\Big\rangle_{t,j}} \ .
\label{eq2.3.1}
\end{equation}
We emphasize that the total passive response per share $R_{i,0}^{(p)}(\tau)$ to be defined later on is not simply the sum of the above two functions $R_{i,0}^{(p, C)}(\tau)$ and $R_{i,0}^{(p, S)}(\tau)$ which measure the passive responses related to the cross-- and self--correlators of trade signs, respectively. Likewise, for the active response per share we define the two contributions by averaging over different $j$, now being the first index,
\begin{equation}
R_{i,0}^{(a, C)}(\tau) \ = \ \frac{\left\langle R_{ji}^{(C)}(\tau)\right\rangle_{j}}{\Big\langle f_j(v_j)\Big\rangle_{t,j} } 
\qquad \textrm{and} \qquad
R_{i,0}^{(a, S)}(\tau) \ = \ \frac{\left\langle R_{ji}^{(S)}(\tau)\right\rangle_{j}}{\Big\langle g_j(v_i)\Big\rangle_{t,j}} \ .
\label{eq2.3.2}
\end{equation}
Again, the total active response $R_{i,0}^{(a)}(\tau)$ to be defined below is not simply the sum. The impact functions of traded volumes are distinguished by the individual stock with the subscript $i$ of the passive response with the superscript $(p)$ or the active response with $(a)$. As the average impact of traded volumes is unrelated to the traded time, we omit the argument $t'$ of the traded volumes. It is worth mentioning that those average impact functions of traded volumes as denominators in Eqs.~\eqref{eq2.3.1} and \eqref{eq2.3.2} are quite different. For the passive response of stock $i$, $\langle f_i(v_i)\rangle_{t}$ and $\langle g_i(v_j)\rangle_{t,j}$ quantify the impacts of traded volumes, respectively, from stock $i$ and stocks $j$ on the price of stock $i$ on average. For the active response of stock $i$, $\langle f_j(v_j)\rangle_{t,j}$ and $\langle g_j(v_i)\rangle_{t,j}$ measure how the traded volumes of stocks $j$ and stock $i$, respectively, influence the average price of stocks $j$.

To further clarify the mechanisms in our model, we now consider it for the following three scenarios. 
\begin{description}  [leftmargin=2.2cm, style=sameline]
\item[ Scenario I] \emph{The cross--impact of trading information from other stocks is very weak, which allows us to set $G_{ij}(\tau)\rightarrow 0$. Therefore, the price cross--response only comes from the cross-correlators of trade signs $\Theta_{ij}(\tau)$.}
\end{description}
\begin{description}  [leftmargin=2.35cm, style=sameline]
\item[ Scenario II] \emph{The cross--correlator of trade signs $\Theta_{ij}(\tau)$ is small enough to be ignored. Therefore, the cross--response only comes from the self--correlator of trade signs, while the cross--impact $G_{ij}(\tau)$ transmitting trading information between stocks is important.}
\end{description}
\begin{description}  [leftmargin=2.5cm, style=sameline]
\item[ Scenario III] \emph{Both the self-- and the cross--correlators of trade signs are responsible for the price cross--response with non--negligible self-- and cross--impacts.}
\end{description}
The average response functions and average impact functions for the three scenarios are discussed in detail in Secs.~\ref{section2.4}, \ref{section2.5} and \ref{section2.6}, respectively. For convenience and to avoid a cumbersome notation, we set the time $t$ at which every trade is executed to zero, $t=0$.

\subsection{Scenario I: Cross--response related to trade sign
  cross--correlators} 
\label{section2.4}

When the cross--impact function approaches zero, the response component $R_{ij}^{(S)}(\tau)$ related to the trade sign self--correlators vanishes, and only $R_{ij}^{(C)}(\tau)$ related to the cross--correlators remains. Hence, the price response across stocks is rooted in the sign cross--correlation. As for the price change, the self--impact function cannot be neglected, although it does not contribute to the correlations across stocks. From Eqs.~\eqref{eq2.2.8}, \eqref{eq2.3.1} and \eqref{eq2.3.2}, the passive and active response functions per share follow as
\begin{equation}
R_{i,0}^{(p)}(\tau)=\sum_{0\leq t<\tau}G_{ii}(\tau-t)\Theta_i^{(p)}(t)+\sum_{t<0}\Big[G_{ii}(\tau-t)-G_{ii}(-t)\Big]\Theta_i^{(a)}(-t) \ ,
\label{eq2.4.1}
\end{equation}
\begin{equation}
R_{i,0}^{(a)}(\tau)=\sum_{0\leq t<\tau}\langle G_{jj}(\tau-t)\rangle_j\Theta_i^{(a)}(t)+\sum_{t<0}\Big[\langle G_{jj}(\tau-t)\rangle_j-\langle G_{jj}(-t)\rangle_j\Big]\Theta_i^{(p)}(-t) \ . 
\label{eq2.4.2}
\end{equation}
When performing averages over stock indices, the trade sign correlators and impact functions are always assumed to be independent of each other. Therefore, the passive and active trade sign cross--correlators, \textit{i.e.} $\Theta_i^{(p)}(\tau)$ and $\Theta_i^{(a)}(\tau)$, appear in Eqs.~\eqref{eq2.4.1} and \eqref{eq2.4.2}.

To facilitate a comparison with the theoretical impact functions resulting from simulations to be discussed in Sec.~\ref{section4}, we transform the average response functions further. In a first step, by substituting $\tau'$ for the time intervals $\tau-t$ and $-t$ in impact functions of Eq.~(\ref{eq2.4.1}) and Eq.~(\ref{eq2.4.2}), the passive and active response functions become
\begin{equation}
R_{i,0}^{(p)}(\tau)= \sum_{\tau'=1}^{\infty}A_{i}^{(p)}(\tau,\tau')G_{ii}(\tau') \ ,
\label{eq2.4.3}
\end{equation}
\begin{equation}
R_{i,0}^{(a)}(\tau)=\sum_{\tau'=1}^{\infty}A_{i}^{(a)}(\tau,\tau')\langle G_{jj}(\tau')\rangle_j \ ,
\label{eq2.4.4}
\end{equation}
where we introduce 
\begin{eqnarray}
A_{i}^{(p)}(\tau,\tau')=\left\{
\begin{array}{lll}
\Theta_i^{(p)}(\tau-\tau')-\Theta_i^{(a)}(\tau')&\ ,\qquad & \mbox{if} \quad 0<\tau'\leq\tau\leq\infty \ , \\
\Theta_i^{(a)}(\tau'-\tau)-\Theta_i^{(a)}(\tau')&\ ,\qquad & \mbox{if} \quad 0<\tau <\tau'\leq \infty \ , 
\end{array}
\right.
\label{eq2.4.5}
\end{eqnarray}
\begin{eqnarray}
A_{i}^{(a)}(\tau,\tau')=\left\{
\begin{array}{lll}
\Theta_i^{(a)}(\tau-\tau')-\Theta_i^{(p)}(\tau')&\ ,\qquad & \mbox{if} \quad 0<\tau'\leq\tau\leq\infty \ , \\
\Theta_i^{(p)}(\tau'-\tau)-\Theta_i^{(p)}(\tau')&\ ,\qquad & \mbox{if} \quad 0<\tau <\tau'\leq \infty \ .
\end{array}
\right.
\label{eq2.4.6}
\end{eqnarray}
Equations~\eqref{eq2.4.5} and \eqref{eq2.4.6} guarantee the positivity of the time lags in the passive and active trade sign correlators. The second step of the transformation is to employ a matrix notation. As we use discretized time, the quantities for different time lags $\tau$ or $\tau'$ can be treated as elements of average response vectors $R_{i,0}^{(p)}$ and $R_{i,0}^{(a)}$, impact vectors $G_{ii}$ and $\langle G_{jj}\rangle_j$, and sign correlation matrices $A_{i}^{(p)}$ and $A_{i}^{(a)}$. We arrive at the rather concise
expressions
\begin{equation}
R_{i,0}^{(p)}=A_{i}^{(p)}G_{ii}
\qquad \textrm{and} \qquad
R_{i,0}^{(a)}=A_{i}^{(a)}\langle G_{jj}\rangle_j\ ,
\label{eq2.4.7}
\end{equation}
which may be inverted,
\begin{equation}
G_{ii}=[A_{i}^{(p)}]^{-1}R_{i,0}^{(p)}
\qquad \textrm{and} \qquad
\langle G_{jj}\rangle_j=[A_{i}^{(a)}]^{-1}R_{i,0}^{(a)}	\ .
\label{eq2.4.8}
\end{equation}
These expressions render it possible to calculate the empirical impact functions from the empirically found responses per share and trade sign correlators. As the above vectors and matrices have infinite dimensions, we use a large cut--off $T_\textrm{cut}$ of 3000 seconds for calculations.

\subsection{Scenario II: Cross--response related to trade sign
  self--correlators}
\label{section2.5}

According to Eq.~\eqref{eq2.2.9}, the information propagator $G_{ij}(\tau)$ which transmits the trading information revealed by the self--correlators of trade signs across stocks is crucial in this scenario. Interestingly, the trade sign self--correlator not only relates to the self--response in single stocks~\cite{Bouchaud2004,Lillo2004} but also to the cross--response between stocks. When considering different stocks, we group the impacts of trading information either coming from different stocks or transmitted to different stocks into an individual impact function, \textit{i.e.}  $G_i^{(p)}(\tau)$ for the former and $G_i^{(a)}(\tau)$ for the latter. Here, $G_i^{(p)}(\tau)$ is the price impact of stock $i$ due to all single trades of different stocks, while $G_i^{(a)}(\tau)$ is the impact of a single trade of stock $i$ on the average price of different stocks. We refer to $G_i^{(p)}(\tau)$ and $G_i^{(a)}(\tau)$ as to the passive and active impact function of stock $i$, respectively. Hence, from Eqs.~\eqref{eq2.2.9}, \eqref{eq2.3.1} and \eqref{eq2.3.2}, we calculate 
\begin{equation}
R_{i,0}^{(p)}(\tau)=\sum_{0\leq t<\tau}G_i^{(p)}(\tau-t)\left\langle\Theta_{jj}(t)\right\rangle_{j}+\sum_{t<0}\Big[G_i^{(p)}(\tau-t)-G_i^{(p)}(-t)\Big] \left\langle\Theta_{jj}(-t)\right\rangle_{j} \ ,
\label{eq2.5.1}
\end{equation}
\begin{equation}
R_{i,0}^{(a)}(\tau)=\sum_{0\leq t<\tau}G_i^{(a)}(\tau-t)\Theta_{ii}(t)+\sum_{t<0}\Big[G_i^{(a)}(\tau-t)-G_i^{(a)}(-t)\Big]\Theta_{ii}(-t) \ ,
\label{eq2.5.2}
\end{equation}
as the passive and active response functions per share.

To obtain the empirical impact functions, transformations similar to the ones from Eq.~\eqref{eq2.4.3} to Eq.~\eqref{eq2.4.8} are carried out. This yields
\begin{equation}
G_{i}^{(p)}=\left [\langle A_{jj}\rangle_j\right]^{-1} R_{i,0}^{(p)} \ ,
\qquad \textrm{and} \qquad
G_{i}^{(a)}=\left [A_{ii}\right]^{-1}R_{i,0}^{(a)} \ ,
\label{eq2.5.3}
\end{equation}
with the matrix elements 
\begin{eqnarray}
A_{ii}(\tau,\tau')=\left\{
\begin{array}{lll}
\Theta_{ii}(\tau-\tau')-\Theta_{ii}(\tau')&\ ,\qquad & \mbox{if} \quad 0<\tau'\leq\tau\leq\infty \ , \\
\Theta_{ii}(\tau'-\tau)-\Theta_{ii}(\tau')&\ ,\qquad & \mbox{if} \quad 0<\tau <\tau'\leq \infty \ . 
\end{array}
\right.
\label{eq2.5.4}
\end{eqnarray}
The matrix elements $\langle A_{jj}\rangle_j$ are defined analogously. Again, the infinity $\infty$ in Eq.~\eqref{eq2.5.4} will be cut off by a large time $T_\textrm{cut}$ of 3000 seconds for calculations. Therefore, Eqs.~\eqref{eq2.4.8} and \eqref{eq2.5.3} reveal the empirical price impacts depending on the time lag $\tau$.

\subsection{Scenario III: Cross--response related to both correlators}  
\label{section2.6}

We now take into account both response components that were individually studied in Scenarios I and II.  Neither the self-- nor the cross--impacts as propagators of single trades can be neglected, both contribute to the price change. We underline once more that both, the cross--impact and the trade sign cross--correlator, generate responses across different stocks. When taking different stocks into account, the impact functions either becomes the average self--impact functions or enters active and passive response functions. Compared to Scenarios I and II, the average response function here describes the price response to trades between stocks completely, regardless of the complexity and hence depending on numerous parameters. By making use of the average response functions per share obtained in Scenarios I and II, we find the average response functions
\begin{equation}  
R_i^{(p)}(\tau)=R_{i,0}^{(p, C)}(\tau)\big\langle f_i(v_i)\big\rangle_{t} +R_{i,0}^{(p, S)}(\tau)\big\langle g_i(v_j)\big\rangle_{t,j} \ , 	
\label{eq2.6.1}
\end{equation}
\begin{equation}  
R_i^{(a)}(\tau)=R_{i,0}^{(a, C)}(\tau)\big\langle f_j(v_j)\big\rangle_{t,j}+R_{i,0}^{(a, S)}(\tau)\big\langle g_j(v_i)\big\rangle_{t,j} \ .	
\label{eq2.6.2}
\end{equation}
The passive response functions, $R_{i,0}^{(p, C)}(\tau)$ and $R_{i,0}^{(p, S)}(\tau)$ have the same forms as in Eqs.~\eqref{eq2.4.1} and~\eqref{eq2.5.1}, respectively. Similarly, the active response functions $R_{i,0}^{(a, C)}(\tau)$ and $R_{i,0}^{(a, S)}(\tau)$ have the same forms as in Eqs.~\eqref{eq2.4.2} and~\eqref{eq2.5.2}, respectively.

\section{Empirical analysis} 
\label{section3}

We analyze the memory properties of the trade sign correlators and the relation between the responses and the traded volumes. In Sec. \ref{section3.1}, we discuss the data set and introduce some definitions, \textit{e.g.} trade sign and time scale. In Sec. \ref{section3.2}, we check the memory properties of the self-- and cross--correlators of trade signs for 31 stocks. In Sec. \ref{section3.3}, we analyze the impacts of traded volumes for passive and active cross--responses of individual stocks.

\subsection{Data sets and definitions} 
\label{section3.1}

The empirical analysis employs the Trades and Quotes (TAQ) data set. Among all the markets included in TAQ data set, we only use the data from NASDAQ stock market in 2008, since NASDAQ is a purely electronic market. To investigate the average response across different stocks, we choose the first 31 stocks from S$\&$P500 index (see Appendix~\ref{appA}) with the largest average number of daily trades. As in Refs.~\cite{Wang2016a,Wang2016b}, we use the physical time instead of the trading time which is convenient when considering self--responses in single stocks~\cite{Bouchaud2004,Lillo2004}. However, when looking at different stocks which each have their own trading time, we found that the physical time is the better choice. As our data has a one--second resolution, it is only meaningful to define the number of daily trades as none or one per second from 9:40 to 15:50 New York local time on the physical time scale, even though more than one trade or quote can occur in this second. To determine the sign of every trade in the one--second interval, we cannot employ the approach of comparing the trades price with the preceding midpoint price in the best quote~\cite{Lee1991}, since trades and quotes data are listed in two individual files without sufficiently short time stamps to specify the preceding midpoint price of the trade. Instead, we employ our approach~\cite{Wang2016a}. If there are $N(t)$ trades in the time interval labeled by $t$, then the trades are numbered $n=1,...,N(t)$ and the corresponding prices are denoted $S(t;n)$. For two consecutive trades in the interval $t$, the sign of the price change is defined as
\begin{eqnarray}       
\varepsilon(t;n)=\left\{                  
\begin{array}{lll}    
\mathrm{sgn}\bigl(S(t;n)-S(t;n-1)\bigr)  & \ ,  & \qquad \mbox{if} \quad S(t;n)\neq S(t;n-1) \ , \\    
\varepsilon(t;n-1) & \ ,  & \qquad \mbox{otherwise} \ .
\end{array}           
\right.    
\label{eq3.1.1}           
\end{eqnarray}
According to Eq.~\eqref{eq3.1.1}, a buy market order with the trade sign $\varepsilon(t;n)=+1$ is executed if the trade price raises, while a sell market order with $\varepsilon(t;n)=-1$ is executed if the trade price falls. If the trade price is unchanged, the trade sign is set to be the same as the preceding one, because the two consecutive trades with the same trading direction did not exhaust the available volume at the best price. If there are more than one trade in the interval $t$, these trades are aggregated yielding a single trade sign for $t$,
\begin{eqnarray}       
\varepsilon(t)=\left\{                  
\begin{array}{lll}    
\mathrm{sgn}
\left(\sum\limits_{n=1}^{N(t)}\varepsilon(t;n)\right) & \ , \qquad & \mbox{if} \quad N(t)>0 \ , \\    
                                                0 & \ , \qquad & \mbox{if} \quad N(t)= 0 \ ,
\end{array}           
\right.    
\label{eq3.1.2}              
\end{eqnarray}     
The case $N(t)=1$ is included. If the majority of trades in second $t$ was triggered by buy (or sell) market orders, then $\varepsilon(t)=+1$ (or $-1$). If trading did not take place or if there was a balance of buy and sell market orders in the second $t$, the trade sign is set to $\varepsilon(t)=0$.

We only consider those days for a stock pair $(i,j)$ in which trading took place in both stocks. In each such day, the trading time is limited from 9:40 to 15:50 of New York local time, which avoids overnight effects and any artifacts due to opening and closing of the market.

\subsection{Properties of trade sign correlators}  
\label{section3.2}

For both, the self-- and the cross--correlators of trade signs appearing in the response functions in Sec.~\ref{section2}, an empirical check of their memories is called for: For the long--memory sign correlation, a buy (sell) market order is more likely to be followed by other buy (sell) market orders. The price thus changes persistently. For the short--memory sign correlation, a buy (sell) market order is not as often followed by other buy (sell) market orders. Thus, the price is more likely to quickly reverse. Previous studies have found long memory in individual stocks, making the trade sign self--correlator slowly decay in a slow power--law fashion~\cite{Bouchaud2004,Lillo2004,Bouchaud2009}. One way to characterize the long--memory process~\cite{Beran1994} is to use the covariance function $Y(\tau)$. In the present case we may identify this object with the trade--sign correlator. In general, the process under consideration has long memory, if in the limit $\tau \rightarrow \infty$, the covariance function has the form
\begin{equation}
Y(\tau)\sim\tau^{-\gamma}L(\tau) \ ,
\label{eq3.2.1}
\end{equation}
where $0<\gamma<1$. The function $L(\tau)$ has to be slowly varying 
at $\tau\to\infty$~\cite{Embrechts1999}, implying
\begin{equation}
\lim_{\tau \rightarrow \infty}\frac{L(\alpha\tau)}{L(\tau)}=1\ , \quad\textrm{for all $\alpha$} \ .
\label{eq3.2.2}
\end{equation}
This asymptotic characterization ignores the correlation at any smaller time lag. The exponent $\gamma$ determines the rate of decay of the correlation rather than their absolute size and thus also whether the integrated correlation function remains finite. Even a small correlation can generate a long--memory process, characterized by the exponent $\gamma$. The smaller $\gamma$, the longer the memory. In financial markets, the exponent $\gamma$ is often measured via a power--law function of the trade sign correlator,
\begin{equation}
\Theta_{ij} (\tau)\simeq \frac{\vartheta_{ij}}{\tau^{\gamma}} \quad
                 \textrm{for large} \, \tau \ .
\label{eq3.2.3}
\end{equation}
The constant $\vartheta_{ij}$ as well as the exponent $\gamma$ are fit parameters~\cite{Wang2016a,Wang2016b}. In the case $i=j$, the above function is the trade sign self--correlator, while for $i\neq j$ it is the cross--correlator. A more refined functional dependence is not needed, as we are only interested in the long--memory properties.

For the sign self--correlators on the trading time scale, Lillo and Farmer (LF)~\cite{Lillo2004} found $\gamma=0.6$ by analyzing 20 highly capitalized stocks traded in the London Stock Exchange. Bouchaud \textit{et al.}~\cite{Bouchaud2009} measured a value of $\gamma$ ranging from 0.2 to 0.7 for the Paris Stock Exchange, \textit{e.g.} $\gamma\approx0.2$ for France--Telecom, and $\gamma\approx0.67$ for Total. Here, we will work out the sign self--correlations on the physical time scale of 31 individual stocks. The results are listed in Appendix~\ref{appA}. In these stocks, $71\%$ show long--memory with $\gamma<1$, and the rest, $29\%$, show short--memory with $\gamma\geq1$. It is not surprising to find short--memory self--correlators in individual stocks. Similar results were obtained in Ref.~\cite{Eisler2012}. The short--memory is due to a balance of long--memory positive and negative correlations. This might also be an explanation for our findings for the self--correlations. However, the positive correlation dominates in the first 10000 seconds. Thus, we tend more to the our explanation put forward in Ref.~\cite{Wang2016a}. In the one--second time intervals, several trades with the same trading direction may occur, they are aggregated to yield one trade sign. As only the net effect of these trades matters, not their individual effects, they have the same overall effect as if just one trade occurred in this one--second interval. We also work out the average self--correlation of trade signs $\left\langle \Theta_{ii}(\tau)\right\rangle_i$. The 31 stocks considered exhibit a long--memory with $\gamma=0.87$, unaffected by the short--memory of a small part of stocks.

In previous analyses~\cite{Wang2016a,Wang2016b}, we provided considerable evidence that the short--memory of sign cross-correlation is converted into long--memory when averaging across the market. We are thus led to check the memory property for the average cross--correlators of trade signs for each individual stocks across other 30 stocks. We find that the passive cross--correlator $\Theta_i^{(p)}$ of $77.4\%$ of the stocks show long--memory with the $\gamma$ ranging from 0.62 to 1, and the active cross--correlator $\Theta_i^{(a)}$ of all stocks exhibits long--memory with $\gamma$ ranging from 0.75 to 1, see Appendix~\ref{appA}.

\subsection{Impacts of traded volumes}  
\label{section3.3}

According to Eqs~\eqref{eq2.2.8} and \eqref{eq2.2.9}, the traded volumes contribute to the price response. We assume that the impact of traded volumes is independent of the time lag $\tau$. In previous studies, Lillo, Farmer and Mantegna~\cite{Lillo2003} have shown that the impact can be described by a concave function on the trading time scale. More specifically, it is a power law, $R(v)\sim v^{\delta}$. Similar studies have been put forward~\cite{Kempf1999,Plerou2002} for time--aggregated volumes. In the study of Potters \textit{et al.}~\cite{Potters2003}, who analyzed stocks traded at the Pairs Bourse and NASDAQ, also on trading time scale, a logarithmic impact, $R(v)\sim \log(v)$, was found. To the best of our knowledge, studies of the price impact of traded volumes on the physical time scale are so far lacking, but will be provided here. For later comparison, we first work out the impact of traded volumes in individual stocks on the physical time scale. We refer to the aggregation of all traded volumes in a one--second interval as traded volume. To put all stocks on roughly the same footing, we normalize the traded volume of each stock at time $t$ by dividing its average traded volume in that year, 2008,
\begin{equation}
v_i(t)=\frac{T\sum_{n=1}^{N(t)}v_i(t;n)}{\sum_{t=1}^{T}\sum_{n=1}^{N(t)}v_i(t;n)} \ ,
\label{eq3.3.1}
\end{equation}
where $T$ denotes the total trading time, \textit{i.e.} the days of trading in both stocks of a pair multiplied by 22200 seconds in each day of 2008. By binning the traded volumes, we obtain the price response as a function of the traded volumes.

\begin{figure}[tbp]
  \begin{center}
    \includegraphics[width=0.7\textwidth]{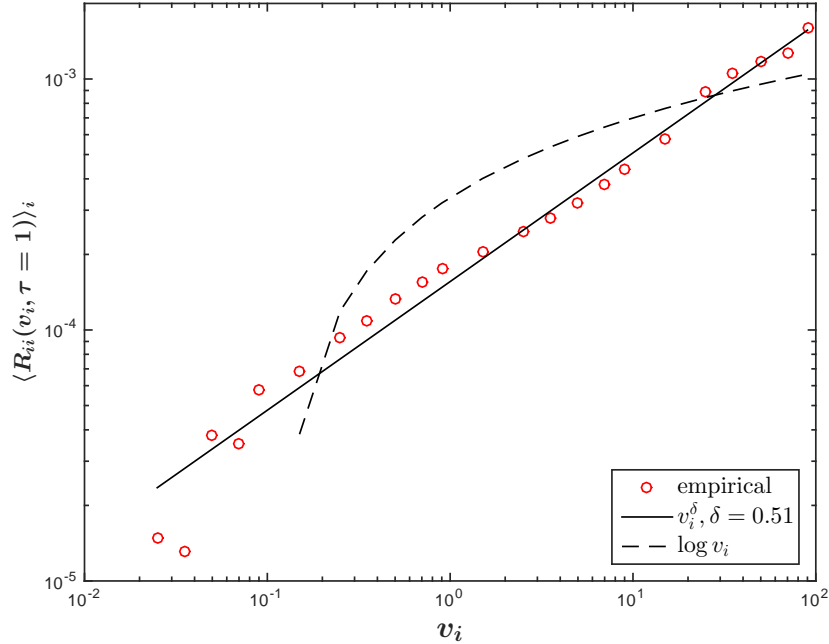} 
  \end{center}
   \setlength{\abovecaptionskip}{-0.3cm} 
\caption{The average self--response over 31 stocks listed in Appendix~\ref{appA} versus traded volumes at time lag $\tau=1$ on a doubly logarithmic scale. The open circles represent the empirical results, the solid line represents the power--law fit, and the dash line represents the natural logarithmical fit.}
 \label{Fig.1}
\end{figure}

Figure~\ref{Fig.1} shows the relation between the price self--response and the traded volumes $\left\langle R_{ii}(v_i,\tau=1) \right\rangle_i$ for individual stocks at time lag $\tau=1$, averaged over the 31 stocks listed in Appendix~\ref{appA}. The fit of power--law and logarithm functions to the empirical results indicates the relation is more in line with the former, $R(v)\sim v^{\delta}$ with an exponent $\delta=0.51$. The exponent value is consistent with previous studies on the trading time scale. Lillo, Farmer and Mantegna~\cite{Lillo2003} found $\delta\sim 0.5$ for small traded volumes and $\delta\sim 0.2$ for large traded volumes in the stocks from New York Stock Exchange in 1995. In another study of Lillo and Farmer, $\delta=0.3$ resulted for Vodafone~\cite{Lillo2004}, one of the five highly capitalized stocks in the London Stock Exchange. The $\delta=0.51$ in our case strongly corroborates the square--root impact function of traded volumes in single stocks, as found in
Refs.~\cite{Torre1997,Gabaix2003,Plerou2004}.

\begin{figure}[htbp]
  \begin{center}
    \includegraphics[width=0.85\textwidth]{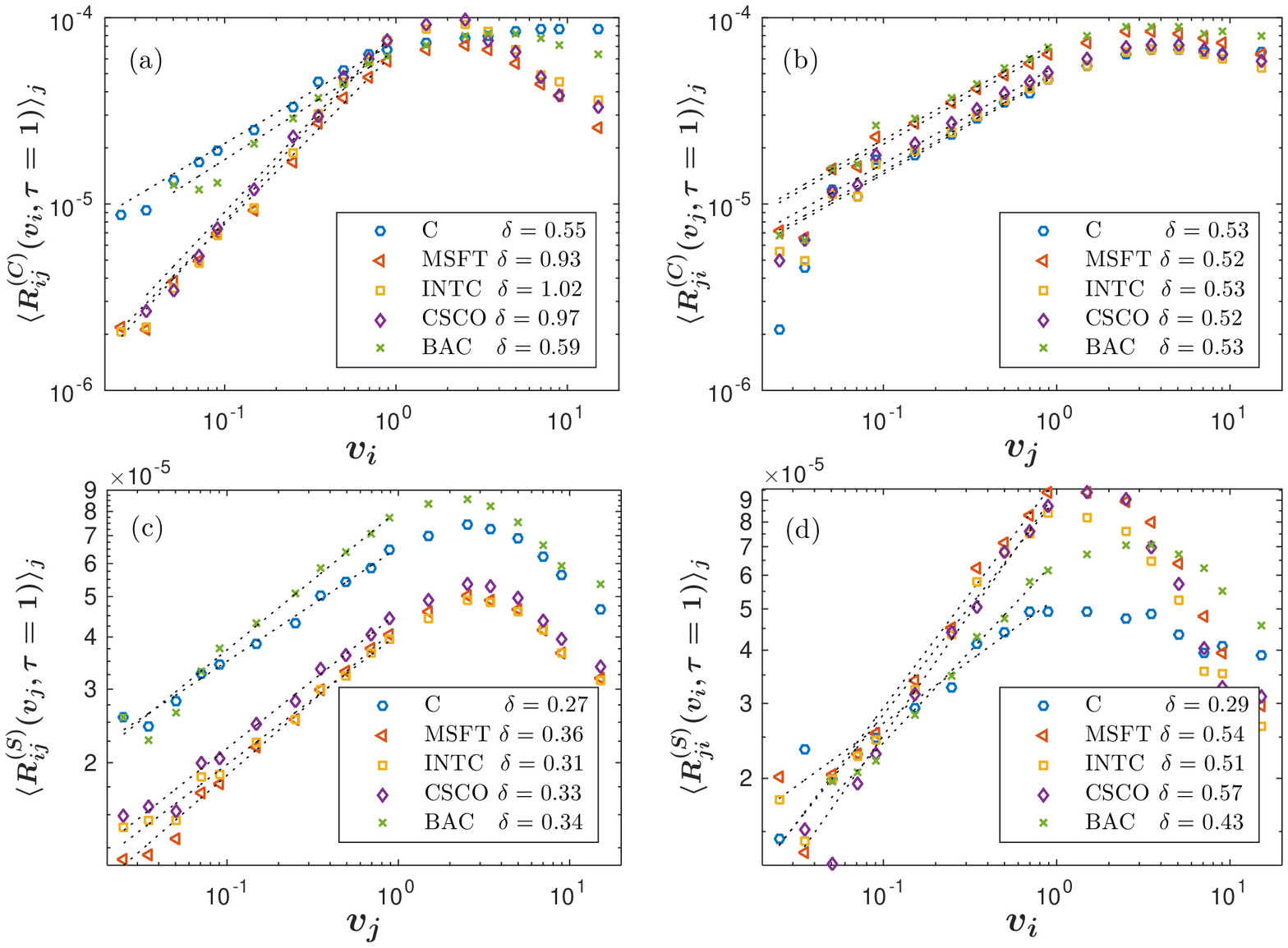} 
  \end{center}
   \setlength{\abovecaptionskip}{-0.2cm} 
\caption{The dependence of the average cross--responses on the traded volumes at time lag $\tau=1$ on a doubly logarithmic scale. (a) The passive response $\langle R_{ij}^{(C)}(v_i,\tau=1)\rangle_j$ of stock $i$ for Scenario I. (b) The active response $\langle R_{ji}^{(C)}(v_j,\tau=1)\rangle_j$ of stock $i$ for Scenario I. (c) The passive response $\langle R_{ij}^{(S)}(v_j,\tau=1)\rangle_j$ of stock $i$ for Scenario II. (d) The active response $\langle R_{ji}^{(S)}(v_i,\tau=1)\rangle_j$ of stock $i$ for Scenario II. Here, the stocks $i$ are C, MSFT, INTC, CSCO and BAC, respectively. The other 30 stocks $j$ are listed in Appendix~\ref{appA}. The markers represent the empirical results and the dot lines represent the power--law fits with the exponents $\delta$. }
 \label{Fig.2}
\end{figure}
Figure~\ref{Fig.2} displays the dependences of average cross--responses for different stocks $i$ on the traded volumes at $\tau=1$. The stocks $i$ are C, MSFT, INTC, CSCO and BAC, the first five stocks with the largest average daily traded volumes among all stocks we studied. Here, four different dependences are discussed: How does the passive cross--responses $\langle R_{ij}^{(C)}(\tau)\rangle_j$ and $\langle R_{ij}^{(S)}(\tau)\rangle_j$ of the stock $i$ depend on average on the traded volumes of stock $i$ itself (Scenario I, see Fig.~\ref{Fig.2} a) and of the other stocks $j$ (Scenario II, see Fig.~\ref{Fig.2} c), respectively? --- How does the active cross--responses of stock $i$, \textit{i.e.} $\langle R_{ji}^{(C)}(\tau)\rangle_j$ and $\langle R_{ji}^{(S)}(\tau)\rangle_j$ of stock $i$ depend on average on the traded volumes of other stocks $j$ (Scenario I, see Fig.~\ref{Fig.2} b) and of stock $i$ itself (Scenario II, see Fig.~\ref{Fig.2} d), respectively? --- In contrast to the average self--response versus traded volumes in Fig.~\ref{Fig.1}, the average cross--responses raise for small traded volumes but decay for large ones. The nonlinear dependence complicates the impact function of traded volumes, even though the average cross--response as well as the average self--response of each stock $i$ for traded volumes smaller than their average can be fitted by a power law.

\begin{figure}[htbp]
  \begin{center}
    \includegraphics[width=0.7\textwidth]{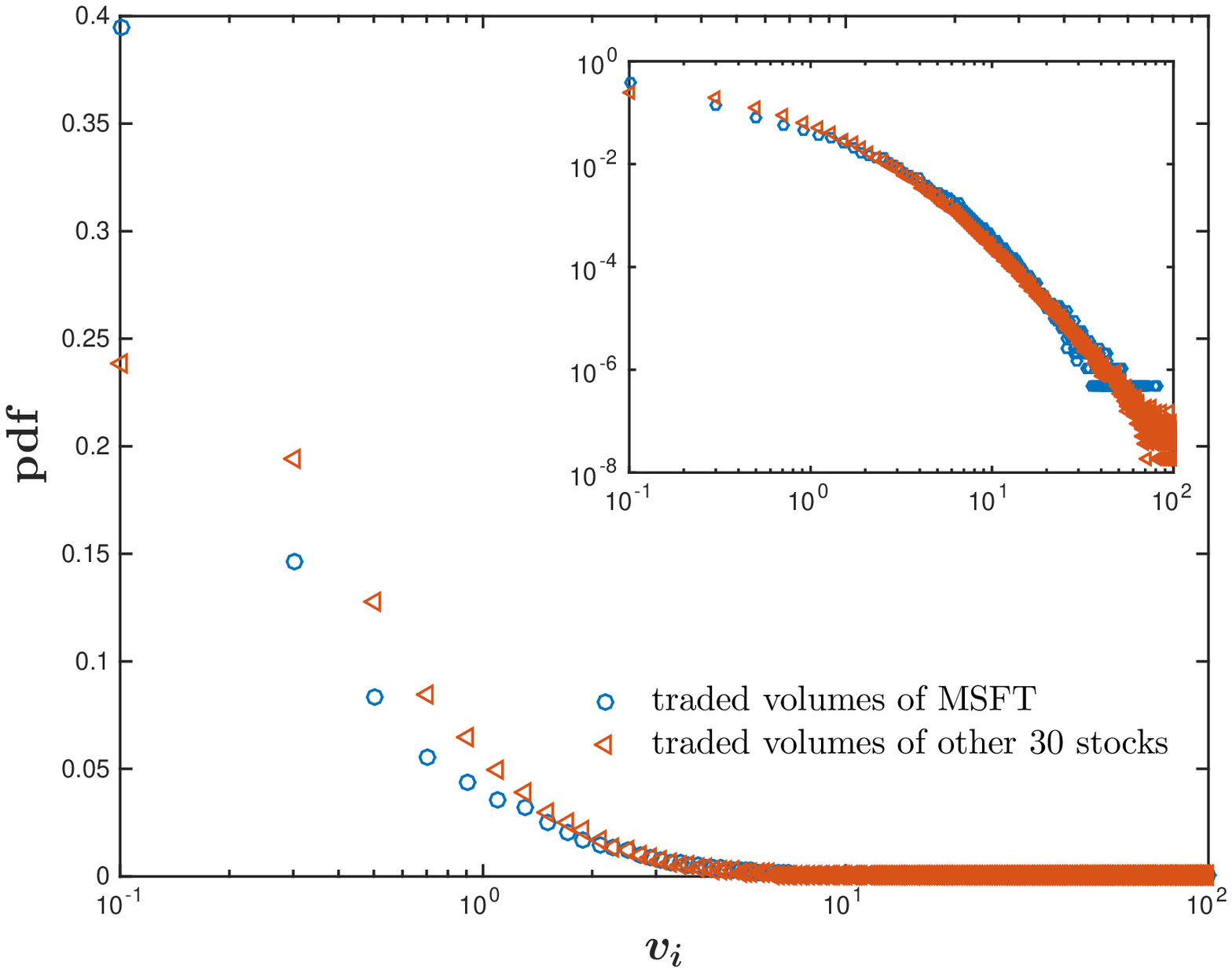} 
  \end{center}
  \setlength{\abovecaptionskip}{-0.3cm} 
\caption{The probability density distribution of traded volumes for MSFT and other 30 stocks on a logarithmic scale. MSFT and the other 30 stocks are listed in Appendix~\ref{appA}. The insert is the probability density distribution of traded volumes on a doubly logarithmic scale.}
 \label{Fig.3}
\end{figure}

To make the analysis feasible, we focus on the region of traded volumes which affect the average cross--response considerably. As an example, we show in Fig.~\ref{Fig.3} the probability density distributions of traded volumes for MSFT and for other 30 stocks paired with MSFT. We conclude that the small volumes are traded frequently, while the large volumes are not. Not surprisingly, it is easier to analyze the self-- or cross--correlators of trade signs in frequently traded stocks. As the power--law dependence in Fig.~\ref{Fig.2} levels off around the average of traded volumes, the ratios of the number of trades with volumes smaller than their average among the total trades are counted. It reaches 73$\%$ of the traded volume smaller than the average for MSFT and to 71$\%$ for the other 30 stocks paired with MSFT. The high proportions narrow down our study to trades below the average traded volume, described well by the power law instead of a complex nonlinear relation.

\begin{figure}[htbp]
  \begin{center}
    \includegraphics[width=0.9\textwidth]{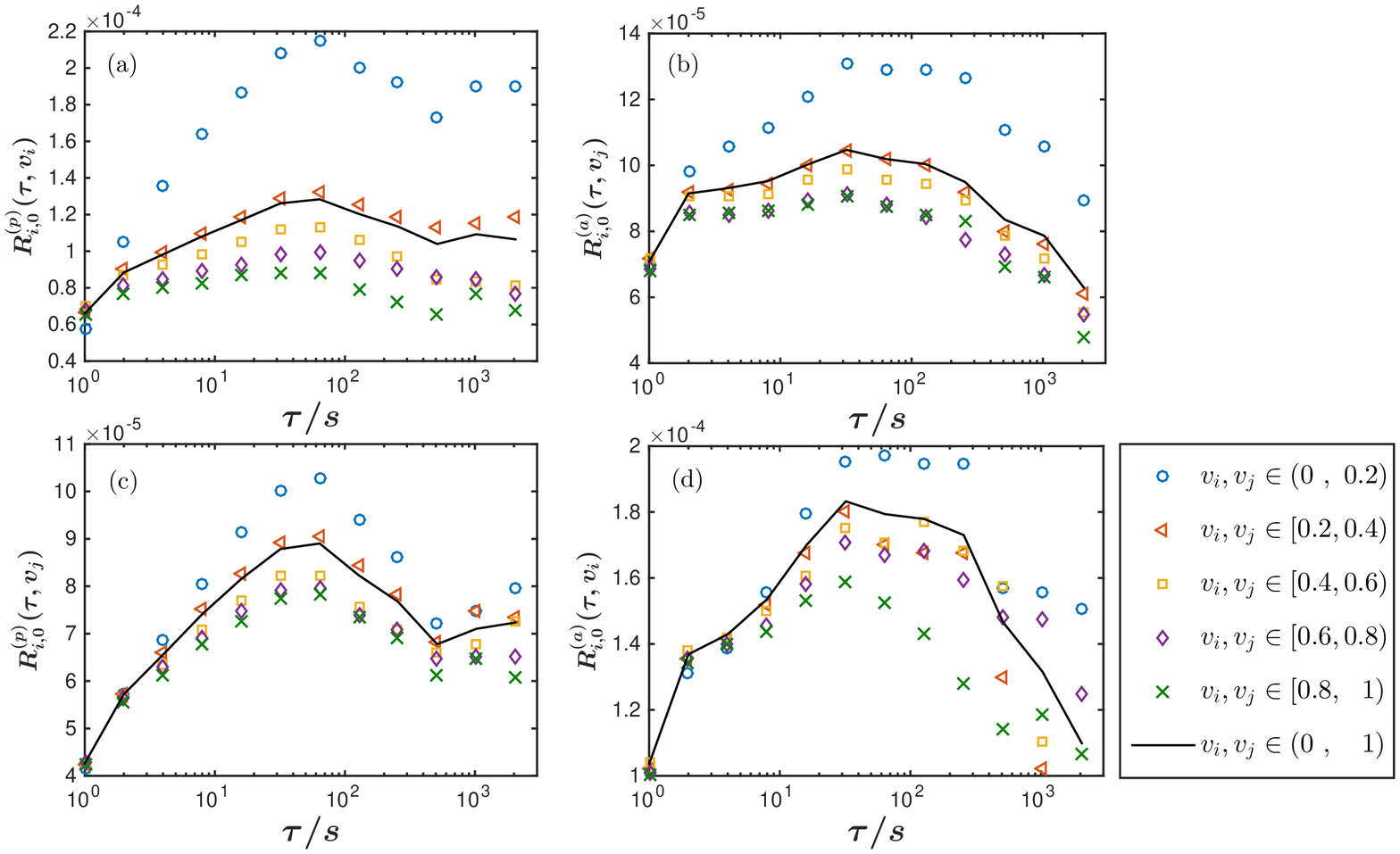} 
  \end{center}
  \setlength{\abovecaptionskip}{-0.2cm} 
\caption{The dependence of average responses per share on the time lag $\tau$ in each region of traded volumes on a logarithmic scale.  (a) The passive responses per share $R_{i,0}^{(p)}(\tau, v_i)$ of stock $i$ in the regions of traded volumes $v_i$ for Scenario I. (b) The active responses per share $R_{i,0}^{(a)}(\tau, v_j)$ of stock $i$ in the regions of traded volumes $v_j$ of other stocks $j$ for Scenario I. (c) The passive responses per share $R_{i,0}^{(p)}(\tau, v_j)$ of stock $i$ in the regions of the traded volumes $v_j$ of other stocks $j$ for Scenario II. (d) The active responses per share $R_{i,0}^{(a)}(\tau, v_i)$ of stock $i$ in the regions of traded volumes $v_i$ for Scenario II. The regions of  $v_i$ and $v_j$ are (0, 0.2), [0.2, 0.4), [0.4, 0.6), [0.6, 0.8), [0.8, 1), and (0, 1), respectively. Here, the stock $i$ is MSFT and the other stocks $j$ are listed in Appendix~\ref{appA}. }
 \label{Fig.4}
\end{figure}
In the power--law function of traded volumes, the exponent $\delta$ indicates the strength of price impact. As visible in Fig.~\ref{Fig.2}, the $\delta$ ranges from 0.27 to 1.02 for the average cross--responses. To be more specific, the comparison of the $\delta$ in Fig.~\ref{Fig.2} a) and c) reveals that the stock price is more likely to be influenced by the traded volumes of the stock itself rather than by those of the other stocks. Put differently, the passive response of stock $i$ depends strongly on the traded volumes of stock $i$ itself, but weakly on the volumes of other stocks $j$. For the active response of stock $i$, the impact of traded volumes from stock $i$ itself differs across different $i$, see Fig.~\ref{Fig.2} d), while this impact from other stocks $j$ basically keeps stable whatever the stock $i$ might be, see Fig.~\ref{Fig.2} b). This is so, because neither the average price changes nor the traded volumes of stocks $j$, over which we average, vary too much for different stocks $i$.

The analyses in Figs.~\ref{Fig.2} and \ref{Fig.3} yield a power--law impact function of traded volumes for the region of volumes smaller than their average. Hence, for passive and active cross-responses of MSFT to the other 30 stocks considered, we arrive at the following approximations of the average impact functions,
\begin{equation}
\langle f_i(v_i)\rangle_{t}\approx 0.28 
\qquad \textrm{and}\qquad 
\langle f_j(v_j)\rangle_{t,j}\approx 0.56 
\qquad \textrm{in Scenario I},
\label{eq3.3.2}
\end{equation} 
\begin{equation}
\langle g_i(v_j)\rangle_{t,j}\approx 0.66
\qquad \textrm{and} \qquad
\langle g_j(v_i)\rangle_{t,j}\approx 0.43 
\qquad \textrm{in Scenario II}.
\label{eq3.3.3}
\end{equation}
which are independent of time lag $\tau$. Dividing those constants leads to the passive and active cross-responses per share, defined in Eqs.~\eqref{eq2.3.1} and \eqref{eq2.3.2}, respectively. Figure~\ref{Fig.4} shows the average cross--responses per share for MSFT versus time lag $\tau$ in each bin of the traded volumes. As seen, small traded volumes have stronger relative responses than large ones. The small traded volumes are more likely due to the fragmentation of large orders, by which the traders try to conceal their trading intention. The consecutive trades of small orders do not only prompt the self--correlator of trade signs in individual stocks~\cite{Bouchaud2004}, but may also lead to the cross--correlator of trade signs between stocks when trying to split two large orders of stocks in the same portfolio.

\section{A construction to quantify price impacts} 
\label{section4}

With numerical simulations and data comparisons, we try to quantify the price impacts. The construction sketched in the sequel should not be viewed as a fit. The number of parameters is forbiddingly large. The purpose of the construction is to show that consistent and reasonable forms of the impact functions lie within the parameter domains. We choose a functional form for the impact and discuss the temporary and permanent impact components in Sec. \ref{section4.1}. We then construct a strategy to determine the parameters for the impact functions in Sec. \ref{section4.2}.

\subsection{Impact function}
\label{section4.1}

\begin{figure}[htbp]
  \begin{center}
    \includegraphics[width=0.8\textwidth]{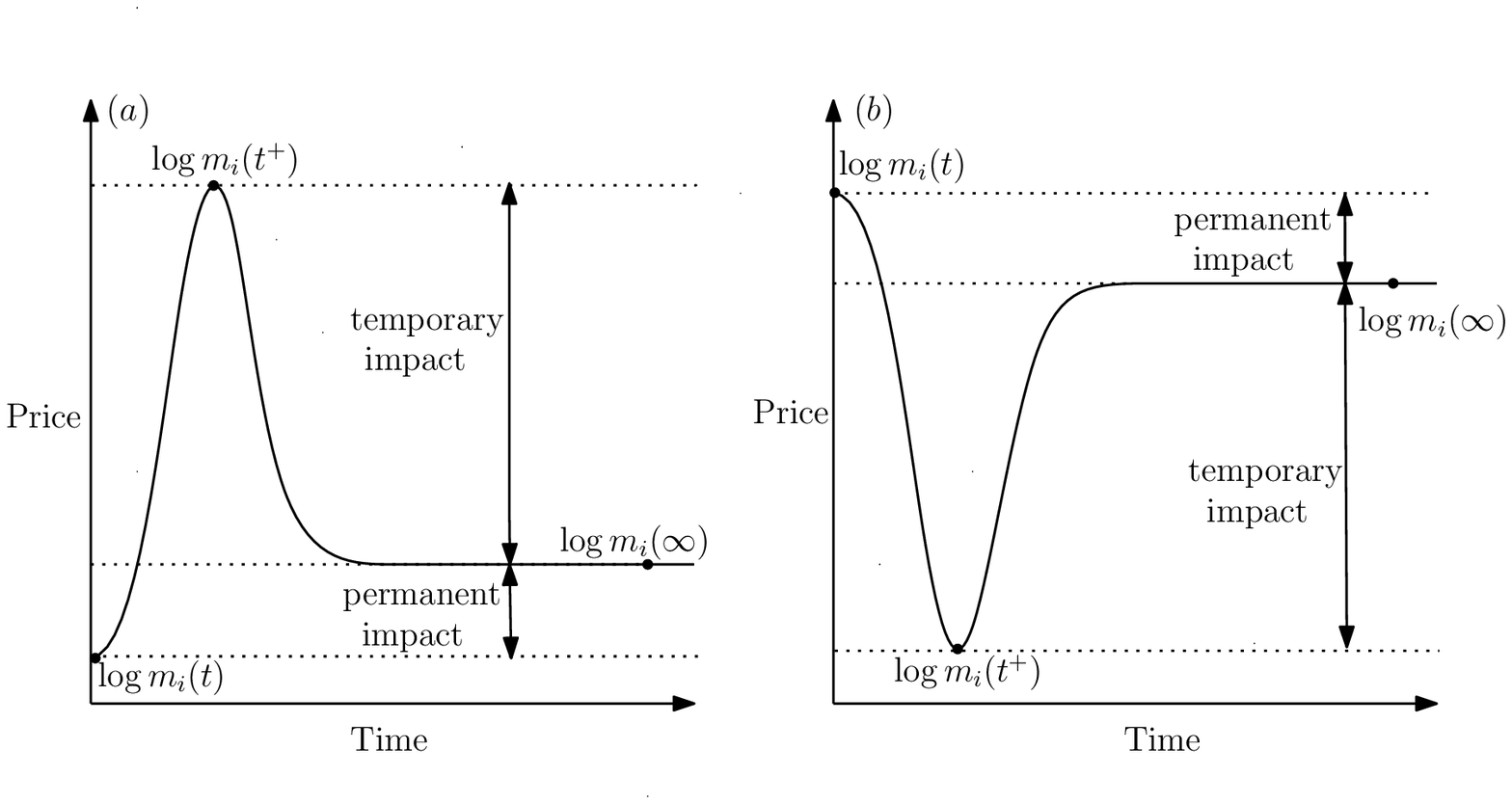} 
  \end{center}
  \setlength{\abovecaptionskip}{-0.3cm} 
\caption{The sketch of the dependence of price impact on the time. (a) the price impact of a buyer--initiated market order. (b) the price impact of a seller--initiated market order.}
 \label{Fig.5}
\end{figure}
The long--memory of the average sign correlations with $\gamma<1$ leads to persistence in the price change. If other factors did not contribute to the price change, the price would be predictable and arbitrage would be possible on long time scales. As this is not only inconsistent with the EMH~\cite{Fama1970} and also in general unrealistic, the price at some larger time lag $\tau$ has to reverse. In a previous study~\cite{Wang2016a}, we argued that the EMH is not valid on short time scales, but restored on longer ones.  The price impact thus has to decay, implying the analogous behavior for the impact function in the response functions.  Mathematically speaking, as the sign correlators $\Theta_{ij}(\tau)$ decay as a power--law function~\eqref{eq3.2.3}, the integrals of the sign correlators over the time lag will increase according to $\tau^{1-\gamma}$ if $\gamma<1$. When considering constant impact functions, the response functions~\eqref{eq2.2.8} and~\eqref{eq2.2.9} will also increase with $\tau^{1-\gamma}$. For $\tau\to\infty$, the response functions will tend to be infinite. A decaying impact function can outmaneuver this divergence. For the small part of stocks with $\gamma>1$, the response functions~\eqref{eq2.2.8} and ~\eqref{eq2.2.9} with $\tau^{1-\gamma}$ will not tend to infinity, and the divergence problem does not emerge.

A single trade can affect the stock price in different ways and with different strengths. It is thus highly unlikely to reverse the price exactly to the previous one while the price impact decays to zero. As the final price differs, in general, from the initial one, the price impact in our model comprises two components, a temporary impact and a permanent impact~\cite{Johnson2010}.  The temporary impact is measured by the difference between the instantaneous price $\log m_i(t^+)$ and the final price $\log m_i(\infty)$, while the permanent impact is measured by the difference between the final price $\log m_i(\infty)$ and the initial price $\log m_i(t)$. Here, the instantaneous price is induced by a buy or sell market order immediately. This is shown in Fig.~\ref{Fig.5}.

The memory properties of the average sign correlator as well as the impact components require the impact function to include two parts,
\begin{equation}
G(\tau)=\frac{\Gamma_0}{\left[1+\left(\frac{\tau}{\tau_0}\right)^2\right]^{\beta/2}}+\Gamma \ ,
\label{eq4.1.1}
\end{equation}
an algebraically decaying term with exponent $\beta$ and with overall strength $\Gamma_0$ over the time scale $\tau_0$ as well as a constant term $\Gamma$. Here, $G(\tau)$ is a general impact function, neither restricted to a special stock nor to an impact type. It stands for the self--impact function $G_{ii}(\tau)$, cross--impact function $G_{ij}(\tau)$, passive impact function $G_{i}^{(p)}(\tau)$ and active impact function $G_{i}^{(a)}(\tau)$, as all of them follow a power-law. In Eq.~\eqref{eq4.1.1}, the decaying term describes the temporary impact component, and converges to the price change with average sign correlations of long--memory.  The constant term provides the permanent impact component, including the possibility of average sign correlations of short--memory. In our price impact model, we use an self-- and a cross--impact function. Hence, for individual stocks, the temporary self--impact comes from the short--run liquidity cost. The difficulty to immediately find willing buyers or sellers induces a price concession from the initial price to the instantaneous price which yields more available volumes for trading~\cite{Holthausen1987, Chan1993}. The permanent self--impact results from private information, which is subsequently incorporated in the new equilibrium price~\cite{Holthausen1987, Chan1993}. The two self--impacts are measured in Ref.~\cite{Almgren2005}. Across stocks, the temporary cross--impact is attributed to the transmission of trading information which, however, is always weakened by competing information. As the strategy traders may benefit much more from this trading information, a permanent cross--impact can result.

According to Eq.~\eqref{eq4.1.1}, the temporary and permanent impact components can be quantified from the impact of a single trade. For instance, if there was a buy market order of stock $i$  at  initial time $t$ with the volume $v_i(t)$. The instantaneous price of stock $i$ for Scenario I is
\begin{eqnarray} \nonumber
\log m_i(t^+)&=&\log m_i(t)+G_{ii}(t^+-t)f_i\big(v_i(t)\big) \\
&=&\log m_i(t)+G_{ii}(0^+)f_i\big(v_i(t)\big)  \ ,
\label{eq4.1.2}
\end{eqnarray}
where the superscript $+$ indicates a time increment smaller than the distance to the next trade.  After the time $\tau$, the restored liquidity due to the new coming limit orders make the price reverse to
\begin{equation}
\log m_i(t+\tau)=\log m_i(t)+G_{ii}(\tau)f_i\big(v_i(t)\big) \ .
\label{eq4.1.3}
\end{equation}
At $\tau\to\infty$ the price change will approach the limit
\begin{eqnarray} \nonumber
\log m_i(\infty)-\log m_i(t)&=&G_{ii}(\infty)f_i\big(v_i(t)\big)\\
	&=&\Gamma f_i\big(v_i(t)\big) \ .
\label{eq4.1.4}
\end{eqnarray}
This is the permanent impact and $\Gamma$ measures the permanent impact per share of stock $i$,
\begin{equation}
\Gamma=\frac{\log m_i(\infty)-\log m_i(t)}{f_i\big(v_i(t)\big)} \ .
\label{eq4.1.5}
\end{equation}
Furthermore, the price reversion of stock $i$ occurs according to
\begin{eqnarray} \nonumber
\log m_i(t^+)-\log m_i(\infty)&=&[G_{ii}(t^+-t)-G_{ii}(\infty)]f_i\big(v_i(t)\big) \\
	&=&\Gamma_0f_i\big(v_i(t)\big) \ ,
\label{eq4.1.6}
\end{eqnarray}
where we assume that the instantaneous impact function $G_{ii}(t^+-t)$ equals the initial one, $G_{ii}(t^+-t)=G_{ii}(0)$. Equation~\eqref{eq4.1.6} states the temporary impact. Correspondingly, $\Gamma_0$ measures the temporary impact per share of the stock $i$,
\begin{equation}
\Gamma_0=\frac{\log m_i(t^+)-\log m_i(\infty)}{f_i\big(v_i(t)\big)} \ .
\label{eq4.1.7}
\end{equation}
Analogously, consider Scenario II and suppose the price change of stock $i$ is triggered by a buy market order of stock $j$. The cross--impact function $G_{ij}(\tau)$ as well as the traded volumes of stock $j$ contribute, such that $\Gamma_0$ and $\Gamma$ measure the temporary and permanent impacts per share of stock $j$, respectively.

\subsection{A construction to fix parameters}  
\label{section4.2}

All impact functions~\eqref{eq4.1.1} have the same form, but the details are encoded in the parameters, \textit{i.e.}, the permanent impact component per share $\Gamma$, the temporary impact component per share $\Gamma_0$, the time scale $\tau_0$ and the rate $\beta$ of the decay. Unfortunately, we cannot determine these impacts directly from empirical data nor observe the latent characteristic of each impact depending on the time lag. To address this problem, we resort to the response functions in Sec.~\ref{section2}, which  comprises the impact functions of time lag, the impact functions of traded volumes and the trade--sign correlators. The last two types of functions can be determined individually from empirical data, because, as analyzed in Sec.~\ref{section3}, the parameters $\vartheta_{ij}$ and $\gamma$ in the trade--sign correlators~\eqref{eq3.2.3} can be obtained by fitting to the empirical correlators, and the average impact functions of traded volumes are approximately constant. Hence, only four parameters in the impact functions~\eqref{eq4.1.1} are to be determined. 

\begin{figure}[t]
  \begin{center}
    \includegraphics[width=0.9\textwidth]{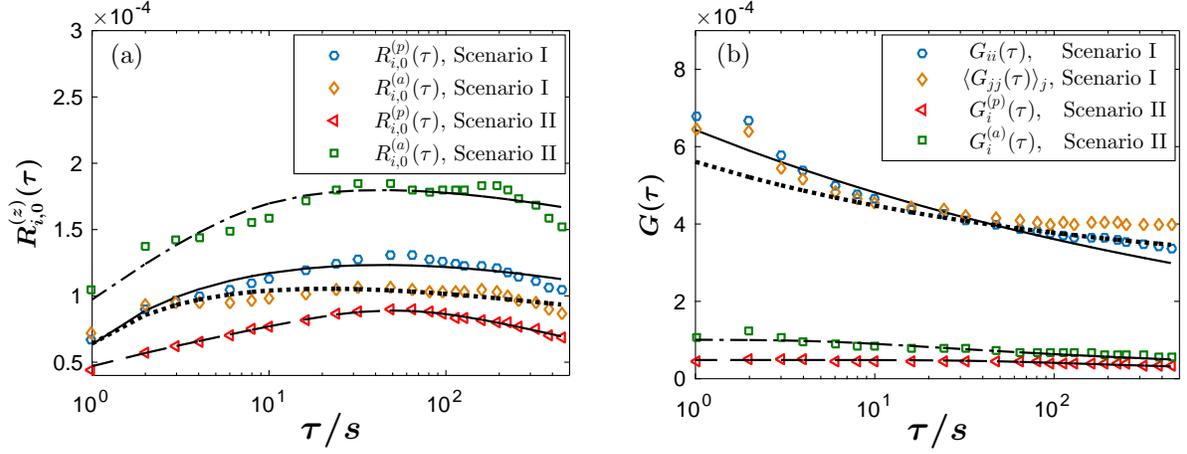} 
  \end{center}
  \setlength{\abovecaptionskip}{-0.2cm} 
\caption{(a) The average cross-responses $R_{i,0}^{(z)}(\tau)$ per share versus the time lag $\tau$ in Scenario I and II, where the superscript $z$ stands for either $a$ or $p$, indicating active and passive cross-responses, respectively. (b) The impact functions $G(\tau)$ versus the time lag $\tau$ in Scenario I and II, where the $G(\tau)$ stands for the self--impact function $G_{ii}(\tau)$ of stock $i$, the average self--impact function $\langle G_{jj}(\tau)\rangle_j$ of other stocks $j$, the passive impact function $G_i^{(p)}(\tau)$ and the active impact function $G_i^{(a)}(\tau)$ of stock $i$, respectively. The stock $i$ is MSFT and the other stocks $j$ are listed in Appendix~\ref{appA}.}
 \label{Fig.6}
\end{figure}

Rather than an exact fit or simulation, we use the following strategy which is merely a construction using consistency arguments to look for appropriate parameters. To begin with, we restrict the response functions to Scenario I and to Scenario II, respectively, where we give initial values that are possible or close to ideal values for the four parameters in the response functions. By adjusting the values of the parameters in $10^6$ iterations to reach a minimal error between the empirical responses per share and the theoretical ones, we find best suited values for the parameters in the Scenarios I and II, respectively. The comparisons of empirical and theoretical results for price responses and impacts are shown in Fig.~\ref{Fig.6}, where the empirical impacts in Scenarios I and II result from Eqs.~\eqref{eq2.4.8} and \eqref{eq2.5.3}, respectively. The fit parameters and the normalized errors $\chi^2$~\cite{Bevington2003} to present the goodness of fits are listed in Table~\ref{table1}.

\begin{table}[b]
\caption{The fit parameters and errors for impact functions and response functions} 
\begin{center}
\begin{tabular}{c@{\hskip 0.1in}c@{\hskip 0.2in}c@{\hskip 0.2in}c@{\hskip 0.2in}c@{\hskip 0.2in}c@{\hskip 0.2in}c@{\hskip 0.2in}c@{\hskip 0.2in}c} 
\hlineB{1.5}
Scenario & response& impact & $\Gamma$  		  &$\Gamma_0$        &$\tau_0$& $\beta$&$\chi^2_{R0}$		&$\chi^2_{G}$		\\
	       & function  & function &$(\times10^{-6})$	  &$(\times10^{-4})$	&[ s ]		&		&$(\times10^{-6})$	&$(\times10^{-6})$	\\
  \hline
\multirow{ 2}{*}{I}&$R_{i,0}^{(p)}(\tau)$ &$G_{ii}(\tau)$		&0.0001	&10.24	&0.02	&0.13	&1.14	&6.93 \\
 			  &$R_{i,0}^{(a)}(\tau)$ &$\langle G_{jj}(\tau)\rangle_j$	&260.90	&20.97	&0.00008	&0.21 	&1.04	&10.74 \\
\hline
\multirow{ 2}{*}{II}&$R_{i,0}^{(p)}(\tau)$ & $G_i^{(p)}(\tau)$ 	&1.58	&0.47	&43.25	&0.18	&0.26	&0.38 \\
 			  &$R_{i,0}^{(a)}(\tau)$ & $G_i^{(a)}(\tau)$ 	&0		&1.01	&5.73	&0.16	&1.90	&1.73 \\
 \hlineB{1.5}
 \label{table1}
  \end{tabular}
\end{center}
\end{table}

\begin{figure}[tb]
  \begin{center}
    \includegraphics[width=0.85\textwidth]{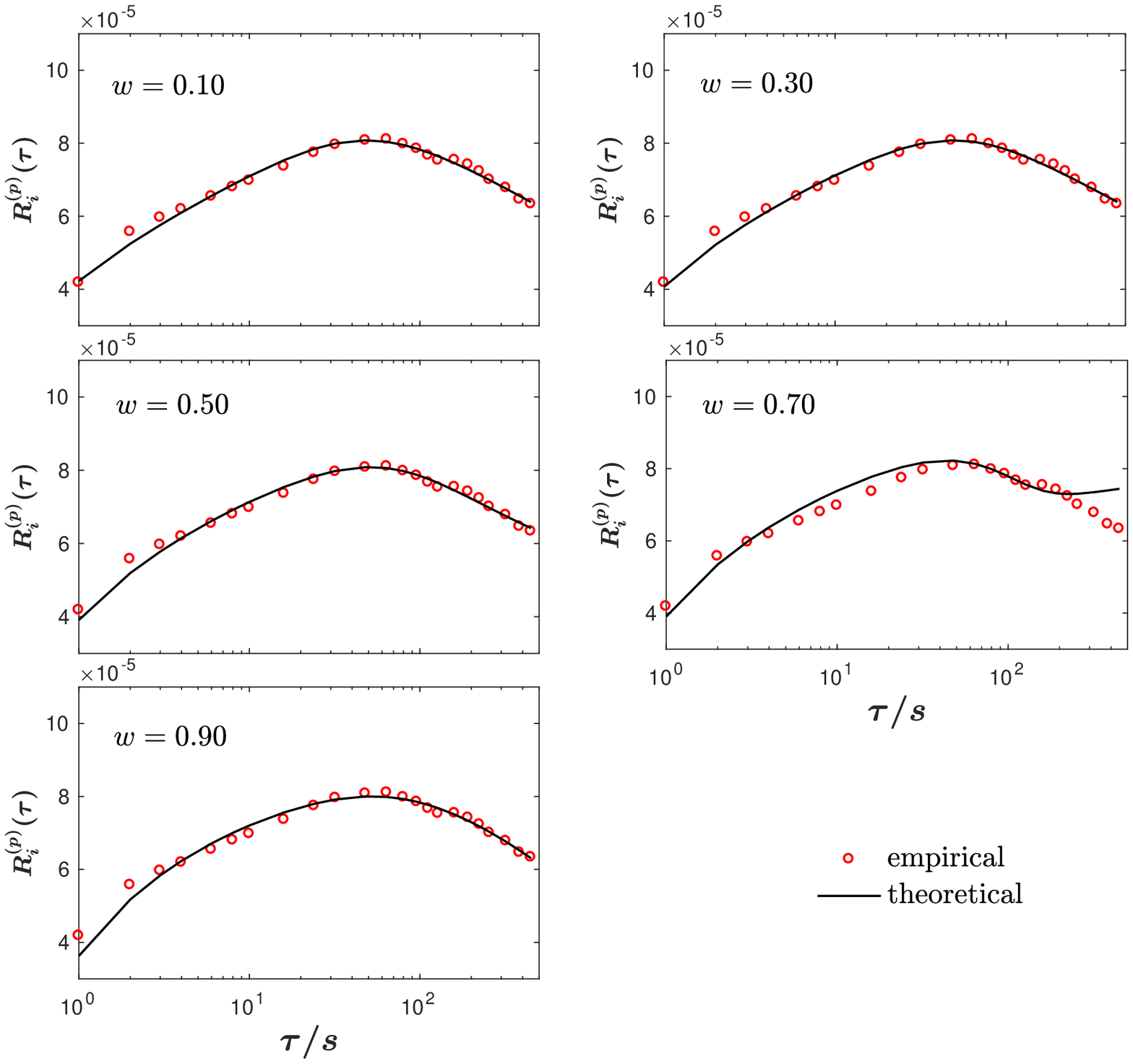} 
  \end{center}
  \setlength{\abovecaptionskip}{-0.2cm} 
\caption{The passive responses versus the time lag $\tau$ in Scenario III with the weights $w=0.10,~0.30,~0.50,~0.70~\textrm{and}~0.90$, respectively.}
 \label{Fig.7}
\end{figure}

\begin{figure}[tb]
  \begin{center}
    \includegraphics[width=0.85\textwidth]{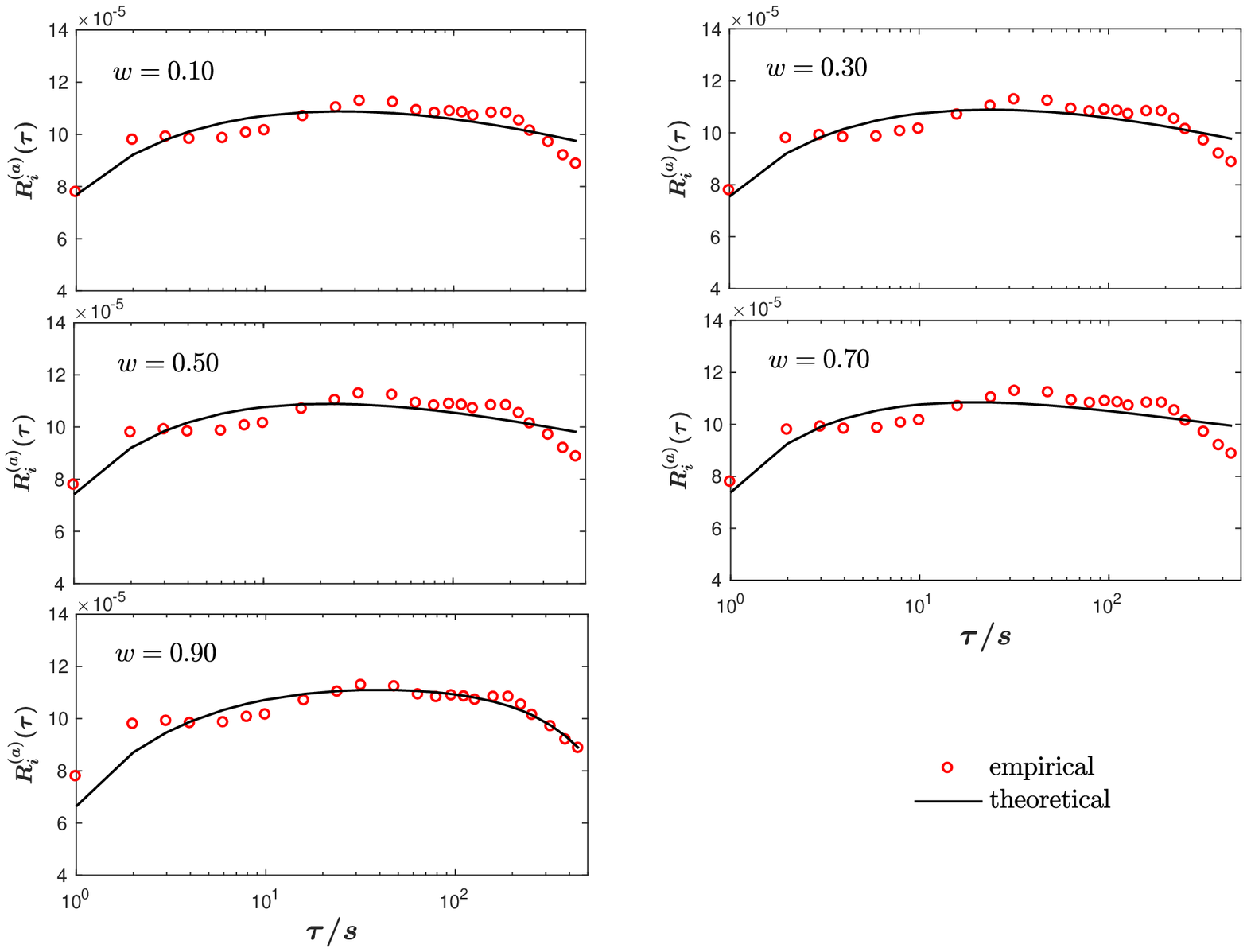} 
  \end{center}
  \setlength{\abovecaptionskip}{-0.2cm} 
\caption{The active responses versus the time lag $\tau$ in Scenario III with the weights $w=0.10,~0.30,~0.50,~0.70~\textrm{and}~0.90$, respectively. }
 \label{Fig.8}
\end{figure}

In a typical market situation, both Scenarios I and II have been accounted for. There are, however, arbitrarily many ways of doing this which all would leave us with too many parameters to be fitted. We thus have no other choice than resorting to an ad hoc method. It turned out that the interpolating ansatz
 \begin{eqnarray}
\label{eq4.2.1}
R_i^{(p)}(\tau)=wR_{i,0}^{(p, C)}\big|_\textrm{I}(\tau)\big\langle f_i(v_i)\big\rangle_{t} +R_{i,0}^{(p, S)} (\tau)\big\langle g_i(v_j)\big\rangle_{t,j} \ ,  \\
R_i^{(a)}(\tau)=wR_{i,0}^{(a, C)}\big|_\textrm{I}(\tau)\big\langle f_j(v_j)\big\rangle_{t,j}+R_{i,0}^{(a, S)} (\tau)\big\langle g_j(v_i)\big\rangle_{t,j} \ ,
\label{eq4.2.2}
\end{eqnarray} 
with a weight $w$ with $0<w<1$ yields good results. Here $R_{i,0}^{(p, C)}\big|_\textrm{I}(\tau)$ and $R_{i,0}^{(a, C)}\big|_\textrm{I}(\tau)$ are the passive and active response per share, respectively, of Scenario I, and $R_{i,0}^{(p, S)} (\tau)$ and $R_{i,0}^{(a, S)} (\tau)$ are the passive and active response per share, respectively, of Scenario III, for which the parameters will be fitted. We proceed as follows: we choose the values $w=0.10,~0.30,~0.50,~0.70~\textrm{and}~0.90$ and for each given $w$ we try to fit the parameters entering $R_{i,0}^{(p, S)} (\tau)$ and $R_{i,0}^{(a, S)} (\tau)$. The results are shown in Fig.~\ref{Fig.7} and \ref{Fig.8} and in Table~\ref{table2}. Once more, we emphasize that the construction to quantify the response functions does not exclude that ansatzes other than \eqref{eq4.2.1} and  \eqref{eq4.2.2} or an altogether different approach might yield comparable or even better results. Our point is only to show that our model is capable of reproducing the data with appropriate parameter values.

\begin{table}[b!]
\caption{The fit parameters and errors for impact functions and response functions in Scenario III} 
\begin{center}
\begin{tabular}{@{\hskip 0.1in}c@{\hskip 0.3in}c@{\hskip 0.3in}c@{\hskip 0.3in}c@{\hskip 0.3in}c@{\hskip 0.3in}c@{\hskip 0.3in}c@{\hskip 0.3in}c@{\hskip 0.1in}} 
\hlineB{1.5}
response &  impact 	  &	$w$	& $\Gamma$  		  &$\Gamma_0$        &$\tau_0$& $\beta$&$\chi^2_{R}$	\\
function   &  function	  &		&				&$(\times10^{-4})$	&	[ s ]	&		&$(\times10^{-6})$	\\
  \hline
	  		&				&0.10         & 0          &0.59         &43.49          &0.18          &0.29\\
          		&				&0.30         & 0          &0.42         &51.55          &0.25          &0.30\\
$R_i^{(p)}(\tau)$&	$G_i^{(p)}(\tau)$&0.50        & 0          &0.25         &70.87          &0.49          &0.35\\
	  		&				&0.70         & 0          &0.11        &270.68         &15.41         &0.98\\
	  		&				&0.90         & 0.03          &-274.43      & 59.68       &-0.0002           &0.45\\
\hline
  
	  		&				&0.10         &0          &3.49          &0.011          &0.15          &1.05\\
	  		&				&0.30         &0          &2.89          &0.008          &0.17          &1.09\\
$R_i^{(a)}(\tau)$&	$G_i^{(a)}(\tau)$&0.50        &0          &2.57          &0.004          &0.19          &1.14\\
	  		&				& 0.70        &0          &2.90          &0.002          &0.28          &1.23\\
	  		&				&0.90         &0.33     &-3292.80   &823.19        &-0.00008          &1.03\\
\hlineB{1.5}
 \label{table2}
  \end{tabular}
  \vspace*{-0.5cm}
\end{center}
\end{table}

The weight determines the proportion of self--impacts in the average cross--responses, giving rise to different features of the price change. To find out the most appropriate weight $w$ and to test how reasonable the parameters for Scenario III are, we need a different function depending on the same parameters. Therefore, in the third step, we approach this issue with the price diffusion function, introduced in Sec.~\ref{section5}.

\section{Relation to correlated diffusion}
\label{section5}

We study how the self-- and cross--impacts are related to the correlated diffusion of prices. We define the diffusion function in Sec.~\ref{section5.1}, and then discuss the time--lag dependence for three stochastic processes modelling the correlated motion of different stock prices in Sec.~\ref{section5.2}. We further analyze the stochastic processes depending on the weight of self--impacts, and compare the empirical and theoretical diffusions to corroborate the choice of parameters in Sec.~\ref{section5.3}. 

\subsection{Price diffusion functions}
\label{section5.1}

We begin with briefly recalling some properties of diffusion in two dimensions.  Consider a particle moving in a flat two--dimensional space with coordinates $(x,y)$. The particle position is changed by random increments $u$ and $v$ in the directions $x$ and $y$, respectively.  We introduce the probability $p(x,y|t)dxdy$ to find the particle in the area element $dxdy$ at time $t$ with the joint probability density $p(x,y|t)$. The partial differential equation for the joint probability density reads
\begin{equation}
\frac{\partial p(x,y|t)}{\partial t}=\frac{\langle u^2\rangle}{2\tau}\frac{\partial^2 p(x,y|t)}{\partial x^2} +\frac{\langle v^2\rangle}{2\tau}\frac{\partial^2 p(x,y|t)}{\partial y^2} +\frac{\langle uv\rangle}{\tau}\frac{\partial}{\partial x}\frac{\partial}{\partial y}p(x,y|t) \ .
\label{eq5.1.1}
\end{equation}
For the convenience of the reader, details are given in Appendix~\ref{appB}. The angular brackets indicate averages over the distribution of $u$ and $v$.  We emphasize the presence of the last term in Eq.~\eqref{eq5.1.1} which is non--zero if the random increments are not independent. Hence, the probability density $p(x,y|t)$ does not factorize. In general, the diffusion equation can be written as~\cite{Gupta2010},
\begin{equation}
\frac{\partial p(\vec{r}|t)}{\partial t}=\nabla\cdot\left(\hat{D}\nabla p(\vec{r}|t)\right) \ , 
\label{eq5.1.2}
\end{equation}
where $\hat{D}$ is a symmetric diffusion tensor in homogeneous and anisotropic media,
\begin{equation}
\hat{D}=\left[\begin{array}{cc}\hat{D}_{xx} & \hat{D}_{xy} \\\hat{D}_{yx} & \hat{D}_{yy}\end{array}\right]  \ .
\label{eq5.1.3}
\end{equation}
This tensor is not space dependent and the two--dimensional diffusion equation becomes
\begin{equation}	
\frac{\partial p(x,y|t)}{\partial t} = \hat{D}_{xx}\frac{\partial^2 p(x,y|t)}{\partial x^2}+\hat{D}_{yy}\frac{\partial^2 p(x,y|t)}{\partial y^2}+2\hat{D}_{xy}\frac{\partial}{\partial x}\frac{\partial}{\partial y}p(x,y|t)  \ ,
\label{eq5.1.4}
\end{equation}
with $\hat{D}_{xy}=\hat{D}_{yx}$. Equations~\eqref{eq5.1.1} and \eqref{eq5.1.4} coincide and allow for the identification
\begin{equation}
\langle u^2\rangle=2\hat{D}_{xx}\tau \ ,
\quad 
\langle v^2\rangle=2\hat{D}_{yy}\tau \ ,
\quad \textrm{and}\quad
\langle uv\rangle=2\hat{D}_{xy}\tau \ .
\label{eq5.1.5}
\end{equation}
In another terminology, the elements of the diffusion tensor $\hat{D}$ are constants in the case of Brownian motion, see details in Appendix~\ref{appB}.

We apply the results~\eqref{eq5.1.5} to the motion of two stocks with indices $i$ and $j$ and obtain the price diffusion function for these two different stocks,
\begin{equation}
D_{ij}(\tau) = \Big\langle r_{ij}(t,\tau) \, r_{ji}(t,\tau)\Big\rangle_{t} \ .
\label{eq5.1.6}
\end{equation}
This diffusion function can be positive, negative or zero. To accumulate statistics, it is helpful to carry out an additional average over the stock index $j$ which defines the quantity
\begin{equation}
\langle D_{i}\rangle (\tau) = 
         \Big\langle r_{ij}(t,\tau) \, r_{ji}(t,\tau)\Big\rangle_{t,j} \ ,
\label{eq5.1.6a}
\end{equation}
where $r_{ij}(t,\tau)$ and $r_{ji}(t,\tau)$ are the logarithmic midpoint price changes of stocks $i$ and $j$, respectively. They can be calculated by Eq.~\eqref{eq2.1.2}. We notice that the diffusion functions in~\eqref{eq5.1.5} read $2\hat{D}_{xx}\tau$ and so on, \textit{i.e.}  they are linear functions in time, while the diffusion coefficients $\hat{D}_{xx}$, etc, are constants. To test the simulated results for all the scenarios, we employ the price diffusion function, which reflects the price fluctuations with time lag $\tau$. For each stock, due to different causes, \textit{i.e.} the short--run liquidity from the stock itself and the trading information from other stocks, the price change contains two components, \textit{i.e.} Eqs.~\eqref{eq2.2.5} and \eqref{eq2.2.6}. Hence, the price diffusion function can be decomposed into four individual sub--functions for different combinations of the components,
\begin{equation}
\langle D_{i}\rangle(\tau)= \langle D_{i}\rangle^{(LL)}(\tau)+\langle D_{i}\rangle^{(II)}(\tau)+\langle D_{i}\rangle^{(LI)}(\tau)+\langle D_{i}\rangle^{(IL)}(\tau) \ .
\label{eq5.1.7}
\end{equation}
In view of Eq.~\eqref{eq2.2.7}, we define the diffusion functions
$\langle D_{i}\rangle^{(XY)}(\tau)$ with $(XY)$ indexed as $(LL)$, $(II)$, $(LI)$,
and $(IL)$ in the following way
\begin{eqnarray}  \nonumber	
\langle D_{i}\rangle^{(LL)}(\tau)&=&\Big\langle r_{ii}^{(L)}(t,\tau)r_{jj}^{(L)}(t,\tau)\Big\rangle_{t,j} \ , \\	\nonumber
\langle D_{i}\rangle^{(II)}(\tau)&=&\Big\langle r_{ij}^{(I)}(t,\tau)r_{ji}^{(I)}(t,\tau)\Big\rangle_{t,j} \ , \\	\nonumber
\langle D_{i}\rangle^{(LI)}(\tau)&=&\Big\langle r_{ii}^{(L)}(t,\tau)r_{ji}^{(I)}(t,\tau)\Big\rangle_{t,j} \ , \\	
\langle D_{i}\rangle^{(IL)}(\tau)&=&\Big\langle r_{ij}^{(I)}(t,\tau)r_{jj}^{(L)}(t,\tau)\Big\rangle_{t,j} \ .		
\label{eq5.1.8}  
\end{eqnarray}
With Eqs.~\eqref{eq2.2.5} and \eqref{eq2.2.6}, we can cast all diffusion functions into to a unified expression,
\begin{eqnarray}\nonumber
\langle D_{i}\rangle^{(XY)}(\tau)&=&\sum_{t\leq t'<t+\tau}G_1(t+\tau-t')G_2(t+\tau-t') \Theta_1(0)V   \\  \nonumber
&+&\quad\sum_{t'<t}\Big[G_1(t+\tau-t')-G_1(t-t')\Big]\Big[G_2(t+\tau-t')-G_2(t-t')\Big] \Theta_1(0)V \\ 
&+&\quad\langle\Delta_{i}\rangle^{(XY)}(\tau)V +\tau D_{\eta}^{(XY)}	  \ ,
\label{eq5.1.9}
\end{eqnarray}
where the numbers $V$ are averages of products of the traded volumes, see Table~\ref{table3}. The noise contributions $D_{\eta}^{(XY)}$, which are assumed to be constants, stem form the random price fluctuations. Moreover, $\langle\Delta_{i}\rangle^{(XY)}(\tau)$ is the contribution induced by the correlation between the impact functions and the sign correlators, it is given by
\begin{eqnarray}\nonumber
\langle\Delta_{i}\rangle^{(XY)}(\tau)
&=&\sum_{t\leq t'<t''<t+\tau}G_1(t+\tau-t')G_2(t+\tau -t'')\Theta_2(t''-t')\\ \nonumber
&+&\sum_{t\leq t''<t'<t+\tau}G_1(t+\tau-t')G_2(t+\tau -t'') \Theta_1(t'-t'')\\ \nonumber
&+&\quad\sum_{t'< t''<t}\Big[G_1(t+\tau-t')-G_1(t-t')\Big]\Big[G_2(t+\tau-t'')-G_2(t-t'')\Big]\Theta_2(t''-t')   \\ \nonumber  
&+&\quad\sum_{t''< t'<t}\Big[G_1(t+\tau-t')-G_1(t-t')\Big]\Big[G_2(t+\tau-t'')-G_2(t-t'')\Big]\Theta_1(t'-t'')    \\ \nonumber  
&+&\quad\sum_{t\leq t''<t+\tau}\sum_{t'<t}\Big[G_1(t+\tau-t')-G_1(t-t')\Big]G_2(t+\tau -t'') \Theta_2(t''-t')    \\ 
&+&\quad\sum_{t\leq t'<t+\tau}\sum_{t''<t}G_1(t+\tau-t')\Big[G_2(t+\tau-t'')-G_2(t-t'')\Big]\Theta_1(t'-t'')	\ . 
\label{eq5.1.10}  
\end{eqnarray}
Table~\ref{table3} summarizes all appearing quantities appearing in the above Eqs.~\eqref{eq5.1.9} and \eqref{eq5.1.10}. Equation~\eqref{eq5.1.7} describes the price diffusion for Scenario III. As for Scenarios I and II, the price diffusions 
\begin{equation}
\langle D_{i}\rangle(\tau)=\left\{\begin{array}{cc} \langle D_{i}\rangle^{(LL)}(\tau) &\qquad (\textrm{Scenario I})\\ \langle D_{i}\rangle^{(II)}(\tau) &\qquad (\textrm{Scenario II}) \end{array}\right .  \ ,
\label{eq5.1.11}
\end{equation}
result from only one component of price change.
\begin{table}[b]
\renewcommand\arraystretch{1.5}
\caption{The quantities in diffusion functions} 
\begin{center}
\begin{tabular}{cccccccc} 
\hlineB{1.5}
$\langle D_{i}\rangle^{(XY)}(\tau)$  & $V$												& $G_1(\tau)$		& $G_2(\tau)$					& $\Theta_1(\tau)$ 		& $\Theta_2(\tau)$ 	& $D_{\eta}^{(XY)}$\\	
\hline
$\langle D_{i}\rangle^{(LL)}(\tau)$	& $\langle f_i(v_i)f_j(v_j)\rangle_{j}\approx 0.179$	& $G_{ii}(\tau)$ 	&$\langle G_{jj}(\tau)\rangle_j$ 		& $\Theta_i^{(p)}(\tau)$ 	& $\Theta_i^{(a)}(\tau)$	& $D_{\eta}^{(LL)}$	\\
$\langle D_{i}\rangle^{(II)}(\tau)$	& $\langle g_i(v_j)g_j(v_i)\rangle_j\approx 0.308$	& $G_i^{(p)}(\tau)$	& $G_i^{(a)}(\tau)$				& $\Theta_i^{(a)}(\tau)$	&$\Theta_i^{(p)}(\tau)$	& $D_{\eta}^{(II)}$		\\
$\langle D_{i}\rangle^{(LI)}(\tau)$	& $\langle f_i(v_i)g_j(v_i)\rangle_j\approx0.208$	& $G_{ii}(\tau)$		& $G_i^{(a)}(\tau)$				& $\Theta_{ii}(\tau)$		& $\Theta_{ii}(\tau)$			& $D_{\eta}^{(LI)}$		\\
$\langle D_{i}\rangle^{(IL)}(\tau)$	&~~$\langle g_i(v_j)f_j(v_j)\rangle_{j} \approx0.416$~~	& ~~$G_i^{(p)}(\tau)$~~	& ~~~$\langle G_{jj}(\tau)\rangle_j$~~		& ~~~$\langle \Theta_{jj}(\tau)\rangle_j$~~ & ~~~$\langle \Theta_{jj}(\tau)\rangle_j$~~	& $D_{\eta}^{(IL)}$	\\
\hlineB{1.5}
 \vspace*{-0.8cm}
 \label{table3}
 \end{tabular}
\end{center}
\end{table}

\subsection{Correlated motion of prices }
\label{section5.2}

In individual stocks, the price stochastic process is often be interpreted as either be normal diffusion, super--diffusion or sub--diffusion~\cite{Havlin2002}. For normal diffusion, the coefficients, such as $\hat{D}_{xx}$ or $\hat{D}_{yy}$ in Eq.~\eqref{eq5.1.5}, are constant. If the price changes persistently, we have super--diffusion, in which, \textit{e.g.}, a high price is more likely to be followed by another high price. If this process continues for a long time, the price is eventually likely to be higher than the initial one. However, if the process persists only for a short time and then reverses, it is said to change anti--persistently, and typically sub--diffusion occurs. A high price is likely to be followed by a low price, and \textit{vice versa}. This can last for a long time.  In mathematical terms, super-- and sub--diffusion are characterized by a non--linear time--dependence of the diffusion function. The three processes can be associated with the Hurst exponent $H$~\cite{Gorski2002}, which is used to measure long--memory process for the auto--correlation of the time series~\cite{Mastromatteo2014,Lillo2004},
\begin{equation}
D_{ii}(\tau)=\Big\langle r_{ii}^2(t,\tau)\Big\rangle_{t} \sim\tau^{2H} \ .
\label{eq5.2.1}
\end{equation}
Here, $H=1/2$ indicates normal diffusion, while $H>1/2$ and $H<1/2$ correspond to super-- and sub--diffusion, respectively~\cite{Bouchaud2004,Mastromatteo2014}.

Across different stocks, we transfer this way of analysis. The diffusion function $\langle D_{i}\rangle(\tau)$ characterizes the correlated motion of the prices. Hence, we introduce an expression analogous to the above one,
\begin{equation}
\langle D_{i}\rangle(\tau)\sim\tau^{2\lambda} 
\label{eq5.2.2}
\end{equation}
with a new exponent $\lambda$, which is not necessarily equal to $H$. This is related to some studies of fractional Brownian motion~\cite{Mandelbrot1968,Perrin2001}.  For an empirical analysis, it is useful to divide out a linear time dependence according to
\begin{equation}
\sqrt{|\langle D_{i}\rangle(\tau)|/\tau}\sim\tau^{\lambda-\frac{1}{2}}  \ .
\label{eq5.2.3}
\end{equation}
As we are here only interested in the time behavior, we use an absolute value to prevent this expression from being imaginary in case of a negative diffusion function. For normal diffusion with $\lambda=1/2$, we obtain the constant diffusion coefficient. If $\lambda>1/2$, the function \eqref{eq5.2.3} is an increasing function of $\tau$. Thus, compared to the normal diffusion, the correlations increase in time and --- if $\langle D_{i}\rangle(\tau)>0$ --- a high price of one stock becomes more likely to be followed by a high price of another stock. This is super--diffusion for correlated stocks. In contrast, if $\lambda<1/2$, the function \eqref{eq5.2.3} decreases with $\tau$ and the correlation decays as compared to the normal diffusion. This is sub--diffusion for correlated stocks.

\subsection{Consistency of our model}
\label{section5.3}

\begin{figure}[tbp]
  \begin{center}
    \includegraphics[width=0.7\textwidth]{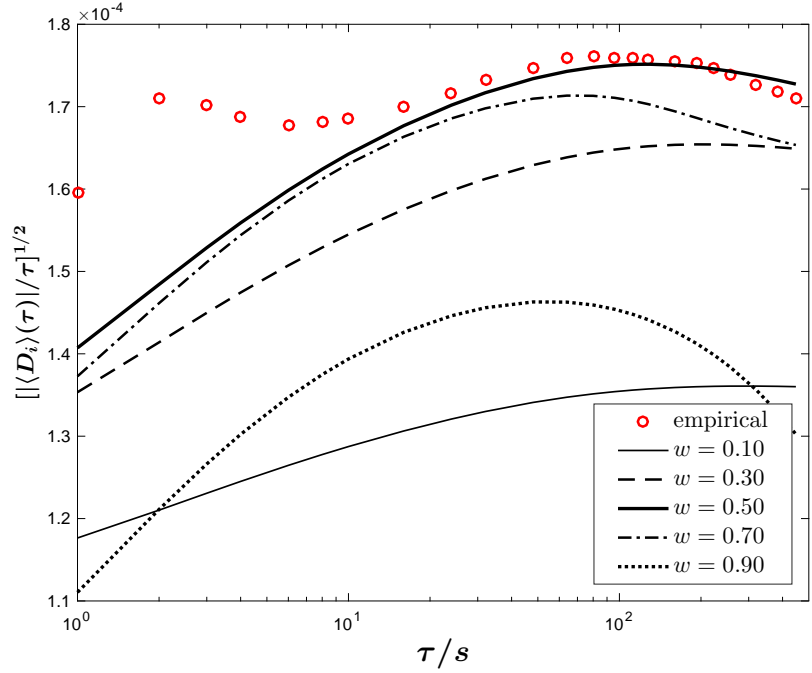} 
     \end{center}
     \vspace*{-0.5cm}
\caption{The comparison of the theoretical results of $[|\langle D_{i}\rangle(\tau)|/\tau]^{1/2}$ in Scenario III, where the weights are $w=0.10,~0.30,~0.50,~0.70~\textrm{and}~0.90$, respectively. These theoretical results are also compared with the empirical result. The random fluctuation is set to $\sum_{(XY)} D_{\eta}^{(XY)}=1\times10^{-8}$ in each case, where $(XY)$ is indexed as $(LL)$, $(II)$, $(LI)$, and $(IL)$ for Scenario III.}
 \label{Fig.9}
\end{figure}

\begin{figure}[htbp]
  \begin{center}
    \includegraphics[width=0.7\textwidth]{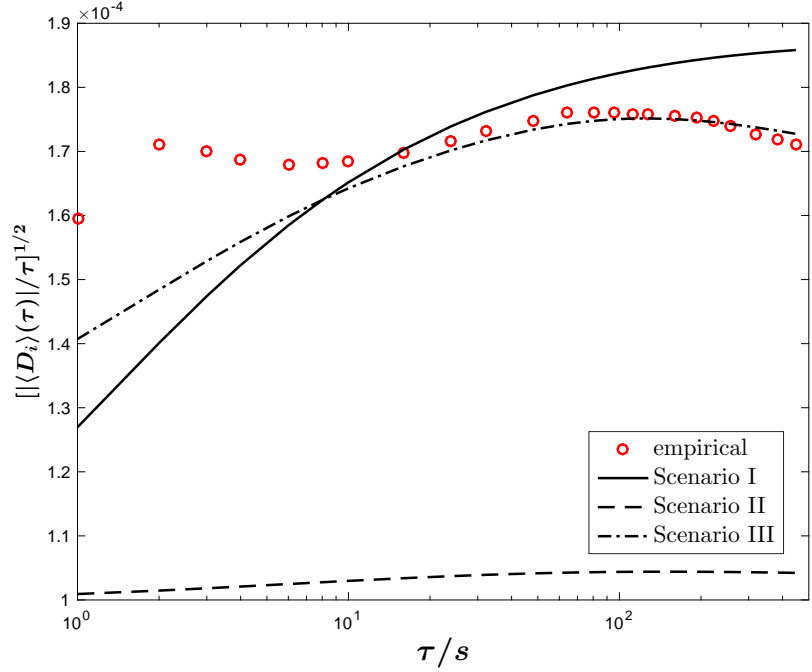} 
     \end{center}
    \vspace*{-0.5cm}
\caption{The comparison of the theoretical results of $[|\langle D_{i}\rangle(\tau)|/\tau]^{1/2}$ in three scenarios, where for Scenario III, we only show the theoretical result with the weight $w=0.5$. These theoretical results are also compared with the empirical result.  The random fluctuation is set to $\sum_{(XY)} D_{\eta}^{(XY)}=1\times10^{-8}$ in each scenario, where $(XY)$ is indexed as $(LL)$ for Scenario I, as $(II)$ for Scenario II, and as $(LL)$, $(II)$, $(LI)$, and $(IL)$ for Scenario III.}
 \label{Fig.10}
\end{figure}

With the parameters in Scenario III, we work out the theoretical price diffusion according to Eq.~\eqref{eq5.1.7}, where the total noise contribution, \textit{i.e.} the sum over the $D_{\eta}^{(XY)}$, is set to $1\times10^{-8}$ for all cases. The comparisons between empirical and theoretical price diffusions are shown in Fig.~\ref{Fig.9}. Theoretical result largely depends on the weight $w$, presenting distinct stochastic processes. For $w<0.50$, due to the small proportion of self--impacts, the cross--impacts dominate in the price change, resulting in a super--diffusive process for the motion of prices in the first 500 seconds. The overestimated cross--impacts fail to reverse the price effectively, and thus provide opportunities of arbitrage, which violates the EMH~\cite{Fama1970}. For $w>0.50$, the process transforms from super--diffusion at first to sub--diffusion later on, as a higher proportion of self--impacts quickly prevents prices from moving in a correlated and persistent manner. The motion of prices in opposite directions opens possible opportunities of arbitrage as well. The consistency between empirical and theoretical diffusions lead us to favor the weight $w=0.50$, especially at larger time lags. In this case, the diffusion coefficient is approximately constant, implying a normal diffusion that the price movement cannot be predicted. Also, we compare the price diffusion in the extreme cases, \textit{i.e.}, Scenarios I and II with only self--impacts or cross--impacts. As shown in Fig.~\ref{Fig.10}, the two scenarios all exhibit super--diffusive behaviors. In contrast, the better agreement for $w=0.5$ in Scenario III lets us conclude that the average cross--responses can indeed be described by the two response components, namely one including the sign self--correlators together with cross--impacts, the other one the sign cross--correlators together with self--impacts. The results are also supported by the small errors $\chi^2_D$ of the price diffusion, listed in Table~\ref{table4}. The above interpretation is also consistent with the line of arguing in Ref.~\cite{Bouchaud2004} where only the self--responses were addressed.

\begin{table}[b]
\begin{center}
\caption{The fit errors of $[|\langle D_{i}\rangle(\tau)|/\tau]^{1/2}$ in three scenarios} 
\begin{tabular}{c@{\hskip 0.25in}c@{\hskip 0.25in}c@{\hskip 0.25in}c@{\hskip 0.25in}c@{\hskip 0.25in}c@{\hskip 0.25in}c@{\hskip 0.25in}c} 
\hlineB{1.5}
  Scenario					&I		&	 II	&\multicolumn{5}{c}{III}		\\
\cline{4-8}
						& 		& 		& $w=0.10$ &$w=0.30$ &$w=0.50$ &$w=0.70$ &$w=0.90$  \\
\hline
$\chi^2_{D} (\times10^{-5})$ 	&0.28        &1.43	&0.85           & 0.31        &0.17         	&0.21          &0.73		\\
\hlineB{1.5}
 \label{table4}
 \end{tabular}
\vspace*{-0.5cm}
\end{center}
\end{table}

 \section{Price impacts of individual stocks}
\label{section6}

\begin{table}[b]
\caption{The parameters of impact functions in Scenario III with $w=0.50$} 
\begin{center}
\begin{tabular}{@{\hskip 0.2in}c@{\hskip 0.3in}c@{\hskip 0.3in}c@{\hskip 0.3in}c@{\hskip 0.3in}c@{\hskip 0.2in}} 
\hlineB{1.5}
 impact 		& $\Gamma$		 &$\Gamma_0$		&$\tau_0$		&$\beta$	\\
 function		&$(\times10^{-10})$	&$(\times10^{-4})$	&	[ s ]		&				\\
  \hline
 $G_{ii}(\tau)$ 		&0.5				&5.12			&0.025		&0.13	\\
  $G_i^{(p)}(\tau)$ 	&0				&0.25			&70.873		&0.49	\\
 $G_i^{(a)}(\tau)$ 	&0				&2.57			&0.004		&0.19	\\
\hlineB{1.5}
 \label{table5}
  \end{tabular}
  \vspace*{-0.5cm}
\end{center}
\end{table}

\begin{figure}[htbp]
  \begin{center}
    \includegraphics[width=0.7\textwidth]{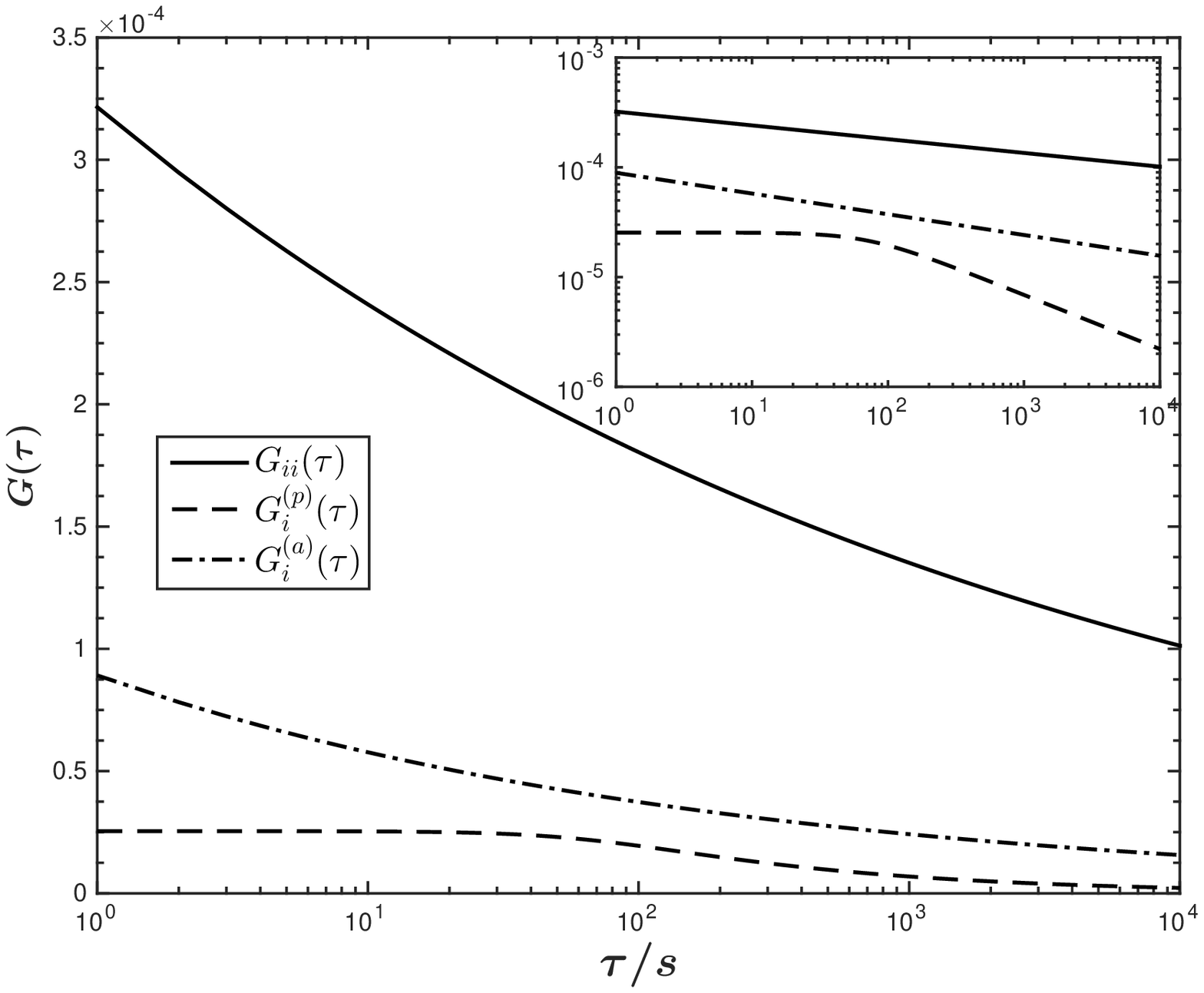} 
     \end{center}
       \setlength{\abovecaptionskip}{-0.3cm} 
\caption{The dependence of price impacts $G_{ii}(\tau)$, $G_i^{(p)}(\tau)$ and $G_i^{(a)}(\tau)$ on the time lag $\tau$ on a logarithmic scale. The price impacts are calculated with the parameters in Scenario III and the weight $w = 0.50$. The insert is the dependence of price impacts on the time lag on a doubly logarithmic scale.}
 \label{Fig.11}
\end{figure}

The price impacts of individual stocks $i$ include the self--impact $G_{ii}(\tau)$ as well as the cross--impacts. For the latter we introduced a passive impact $G_i^{(p)}(\tau)$ and an active impact $G_i^{(a)}(\tau)$. The difficulty to obtain the empirical impacts leads us to describe them with the function~\eqref{eq4.1.1}. Using consistency arguments in Secs.~\ref{section4} and \ref{section5}, we are able to fix the parameters. Thus, the resulting impact function is only a possible one, not excluding others with similar properties. Taking MSFT as an example, with the parameters in Scenario III and the weight $w=0.50$, we are able to provide the price impacts versus time lag, as shown in Fig.~\ref{Fig.11}. The resulting parameters for the impact functions are listed in Table~\ref{table5}.

Table~\ref{table5} shows the temporary and permanent components, $\Gamma_0$ and $\Gamma$ in the self--impact of MSFT, measuring the impact of a single trade of the stock on its own price after a time $\tau$. In our model the self--impact is due to the short--run liquidity in general, however, the existence of the above two components requires more explanation. This is reminiscent of Ref.~\cite{Farmer2007}, where the price impact is separated into a mechanical impact and an informational impact. The mechanical impact of a market order is referred to as the change of future prices without any future change in decision making. The average mechanical impact decays to zero monotonically in time in a power--law fashion, similar to the temporary component in our self--impact. The informational impact is the remainder after the mechanical impact being removed. It grows with time and approaches a constant value, just as the permanent component in our self--impact. Thus, a line of reasoning put forward in Ref.~\cite{Farmer2007} can be partly transferred to our case. As the incoming limit orders following the instantaneous price change offer more liquidity for the market and reverse the trade price towards the previous price, the temporary component is still induced by the short--run liquidity, but the reversed final price is less likely to exactly be the previous one. Therefore, the induced permanent component as well as the informational impact may result from private information. If individual agents possess private information to trade, a price change due to the trading affects will emerge. Other intelligent agents will then adjust their market expectations based partly on the private information and partly on all available public information. The private information is made public via the trade price, visible in the permanent component of the self--impact. However, in contrast to the temporary component of the order of $10^{-4}$, the permanent component is very small, of the order of $10^{-11}$.

For the cross--impact of MSFT, the permanent component is absent. Hence, either passive or active impacts only contain the temporary component. We recall that the passive impact is the price change of an individual stock $i$ induced by single trades of different stocks and the active impact is the average price change of different stocks triggered by a single trade of stock $i$. The cross--impact accounts for the trading information transmitted across stocks. We stress once more that the trading information indicates the trading directions (buy and sell) and trading volumes of other stocks, unrelated to private information, news and so on, which are viewed as competing information. Because of the interference of this competing information, the influence of trading information can neither remain for a long time nor be as strong as the self--impact. Moreover, the difficulty for the traders to distinguish useful and useless trading information makes a permanent impact unlikely.

In addition, although the passive impact shows weak temporary component, it decays very slowly with the decay time scale of $\tau_0=70.873$~s. The decay of the impact at about 70 s is visible in Fig.~\ref{Fig.11}. It is a consequence of the reduction of the noise induced by random trades, as the passive impact extracts the trading information from multiple stocks. In contract, the trading information from only one stock leads to small decay time scales either for self--impact or for active impact.

\section{Conclusions}  
\label{section7}

We put forward a price impact model for the average cross--response functions for individual stocks.  It comprises two impact functions, \textit{i.e.} a self--impact function and a cross--impact function. We introduced and studied three scenarios, namely cross--responses exclusively due to the trade--sign cross--correlators (Scenario I), or to the trade--sign self--correlators (Scenario II), or to both (Scenario III), respectively. Thereby, we managed to greatly reduce the complexity of the problem, and facilitated the determination of the model parameters.

In the empirical analysis we demonstrated that, for most stocks, the self--correlators and average cross--correlators of trade signs have a long memory, which provides a strong support for setting up the impact function of the time lag. The empirical analysis also revealed power--law relations between the average cross--responses and the traded volumes that are smaller than their average. The relations hold regardless of the passive or active cross--responses and regardless of the traded volumes of the impacted or impacting stock. To further explore the parameter space of our model, we defined the average cross--responses per share, which are the average cross--responses divided by the average impact functions of traded volumes.  Our empirical analysis manifests that the smaller the volumes, the larger are the responses per share. This indicates the fragmentation of large orders.

Using the average cross--response functions and the price diffusion functions, we gave a construction to fix the parameters for the impact functions. The results indicate that there are two components present in the response functions. One contains the self--impacts that suppress the amplification effects due to sign cross--correlators, the other one contains the cross--impacts that suppress the amplification effects due to sign self--correlators.

As an example, we studied the price impacts of MSFT. The self--impact includes temporary and permanent components. The temporary component, as a decaying power--law function, is the result of short--run liquidity, while the permanent component, approaching a constant, is due to private information. However, the permanent component is rather small compared to the temporary one. The cross--impacts, separated into an active impact and a passive impact, only contain the temporary component. It comes from the trading information transmitted across stocks. In our study, the trading information is limited to the trading directions, \textit{i.e.} buy and sell, and traded volumes of impacting stocks. The interference of competing information weakens the influence of the trading information. Consequently, the cross--impacts are neither as strong as the self--impact nor persistent permanently.

\section*{Acknowledgements}

We thank M.~Akila, R.~Sch\"afer and Y.~Stepanov for fruitful discussions. One of us (SW) acknowledges financial support from the China Scholarship Council (grant no.~201306890014).

\appendix

\section{Stock information}
\label{appA}

According to the average number of daily trades for each stock in 2008, we select 31 stocks from all available stocks in S$\&$P 500 index of that year. The 31 stocks for your study are listed in Table~\ref{tableA}, where we not only give the information of the economic sector and the average number of daily trades for each stock, but also record the additional information, \textit{e.g.} the average daily traded volume, and the exponents $\gamma$ for the trade sign self--correlator $\Theta_{ii}(\tau)$, for the passive cross--correlator $\Theta_i^{(p)}(\tau)$ and for the active cross--correlator $\Theta_i^{(a)}(\tau)$.

\begin{table}[htbp]
\caption{The average daily trading information and the $\gamma$ values for each stock} 
\begin{center}
\begin{tabular}{l@{\hskip 0.1in}l@{\hskip 0in}c@{\hskip 0.15in}c@{\hskip 0.15in}c@{\hskip 0.15in}c@{\hskip 0.15in}c@{\hskip 0.15in}c} 
\hlineB{1.5}
Symbol	&	~~~~~~~~Sector			&	Average number  &	Average daily  traded	&				& $\gamma$			   &				         \\
\cline{5-7}
		&							&	of daily trades	&  volume ($\times10^6$)	& ~for $\Theta_{ii}(\tau)$~&~for $\Theta_i^{(p)}(\tau)$~&~for $\Theta_i^{(a)}(\tau)$~\\
\hline
AAPL	&	Information Technology		&	13415	&	13.27		&1.36 		&0.71		&0.83\\
JPM 		&	Financials					&	10284	&	12.62		&1.07		&1.06		&0.81\\
XOM  	&	Energy					&	9708		&	7.79			&1.19		&1.50		&0.95\\
BAC  	&	Financials					&	9599		&	18.08		&0.92		&0.89		&0.79\\
WFC 	&	Financials					&	9040		&	12.43		&0.90		&0.88		&0.81\\
MER 	&	Financials					&	8823		&	9.20			&0.97		&0.96		&0.79\\
C 		&	Financials					&	8297		&	30.48		&0.78		&0.68		&0.75\\
QCOM  	&	Information Technology		&	8132		&	8.43			&0.84		&0.87		&0.87\\
MS  		&	Financials					&	7860		&	7.04			&1.00		&0.92		&0.79\\
MSFT  	&	Information Technology		&	7794		&	30.39		&0.70		&0.70		&0.84\\
WMT  	&	Consumer Staples			&	7438		&	6.28			&0.91		&0.94		&0.88\\
CVX  	&	Energy					&	7331		&	3.65			&1.30		&1.51		&0.99\\
GS 		&	Financials					&	7073		&	3.46			&1.23		&0.94		&0.79\\
WB  		&	Financials					&	6856		&	14.41		&0.78		&0.66		&0.75\\
COP  	&	Energy					&	6712		&	3.23			&1.07		&1.24		&0.95\\
CSCO  	&	Information Technology		&	6697		&	22.41		&0.69		&0.67		&0.83\\
CHK  	&	Energy					&	6603		&	4.50			&0.94		&0.92		&0.91\\
INTC  	&	Information Technology		&	6567		&	25.61		&0.65		&0.62		&0.81\\
GE  		&	Industrials					&	6475		&	16.67		&0.78		&0.73		&0.83\\
HAL  	&	Energy					&	6455		&	4.43			&0.90		&0.94		&0.92\\
AMZN 	&	Consumer Discretionary		&	6371		&	3.66			&0.98		&0.86		&0.88\\
FCX 		&	Materials					&	6308		&	3.08			&1.11		&1.11		&0.91\\
T  		&	Telecommunications Services	&	6239		& 	 6.64			&0.83		&0.84		&0.86\\
USB 		&	Financials					&	6078		&	4.49 			&0.88		&0.88		&0.82\\
HPQ  	&	Information Technology		&	6056		&	4.47			&0.86		&0.88		&0.88\\
AXP  	&	Financials					&	6046		&	3.42			&0.96		&0.97		&0.85\\
SLB  	&	Energy					&	5952		&	2.56			&1.20		&1.30		&0.98\\
AIG  		&	Financials					&	5928		&	12.38		&0.82		&0.76		&0.77\\
GILD  	&	Health Care				&	5851		&	3.27			&0.81		&0.92		&0.89\\
PG  		&	Consumer Staples			&	5765		&	3.45			&0.93		&1.12		&0.92\\
ORCL 	&	Information Technology		&	5696		&	14.94		&0.66		&0.65		&0.84\\
\hlineB{1.5}
\label{tableA}
\end{tabular}
\end{center}
\end{table}

\section{Diffusion equation in two dimensions}
\label{appB}

Consider a particle moving in a flat two--dimensional space with coordinates $(x, y)$ at time $t$. After a small and fixed time $\tau$, this particle moves to the position $(x+u, y+v)$. The random increments $u$ and $v$ in the direction $x$ and $y$, respectively, can be positive or negative and satisfies a normalized and symmetric distribution of marginal probability density,
\begin{equation}
\int_{-\infty}^{+\infty}q(u)du=1
\qquad \textrm{and} \qquad
q(-u)=q(u) \ ,
\label{eqB1}
\end{equation}
\begin{equation}
\int_{-\infty}^{+\infty}q(v)dv=1
\qquad \textrm{and} \qquad
q(-v)=q(v) \ .
\label{eqB2}
\end{equation}
Their joint probability density distribution is also normalized to unity between $-\infty$ and $+\infty$,
\begin{equation}
\int_{-\infty}^{+\infty}\int_{-\infty}^{+\infty}q(u,v)dudv=1 \ .
\label{eqB3}
\end{equation}
According to the Eqs~.\eqref{eqB1}--\eqref{eqB3}, the random increments have following properties,
\begin{equation}
\langle uv\rangle= \int_{-\infty}^{+\infty}\int_{-\infty}^{+\infty}uvq(u,v)dudv \ ,
\label{eqB4}
\end{equation}
\begin{eqnarray} 
\langle u^n\rangle =\int_{-\infty}^{+\infty}\int_{-\infty}^{+\infty}u^n q(u,v)dudv  
			    =\int_{-\infty}^{+\infty}u^n q(u)du  \ ,
\label{eqB5}
\end{eqnarray}
\begin{eqnarray}
\langle v^n\rangle =\int_{-\infty}^{+\infty}\int_{-\infty}^{+\infty}v^n q(u,v)dudv  
			    =\int_{-\infty}^{+\infty}v^n q(u)du  \ .
\label{eqB6}
\end{eqnarray}
As the particle moves without any external driving force, the positive and negative increments in each direction have equal probability, which lead to 
\begin{equation}
\langle u\rangle=0
\qquad \textrm{and} \qquad
\langle v\rangle=0 \ .
\label{eqB7}
\end{equation}

We introduce the probability $p(x,y|t)dxdy$ to find the particle in the area element $dxdy$ at time $t$ with the joint probability density $p(x,y|t)$.
Now suppose the particle moves to the position $(x', y')$ at time $t$. After the time increment $\tau$, the probability density to find the particle in the new position $(x, y)$ is
\begin{eqnarray} \nonumber
p(x,y|t+\tau)&=&\int_{-\infty}^{+\infty}\int_{-\infty}^{+\infty}dx'dy'p(x',y'|t)\int_{-\infty}^{+\infty}\int_{-\infty}^{+\infty}dudvq(u,v)\delta(x-(x'+u))\delta(y-(y'+v)) \\
	 	  &=&\int_{-\infty}^{+\infty}\int_{-\infty}^{+\infty}dudvq(u,v)p(x-u,y-v|t)	\ ,
\label{eqB8}
\end{eqnarray}
where the $\delta$ functions $\delta(x-(x'+u))\delta(y-(y'+v))$ as the proper filter fix the $x$ and $y$ to $x'+u$ and $y'+v$, respectively. The random increments $u$ and $v$ during the time increment $\tau$ are assumed to be all very small. Thus the Eq.~(\ref{eqB8}) can be derived as,
\begin{eqnarray} \nonumber
p(x,y|t+\tau)+\tau\frac{\partial p(x,y|t)}{\partial t}	
&=&\int_{-\infty}^{+\infty}\int_{-\infty}^{+\infty}dudvq(u,v)\left\{p(x,y|t)-u\frac{\partial p(x,y|t)}{\partial x}-v\frac{\partial p(x,y|t)}{\partial y} \right.\\    [0.7em]
&+&\left.\frac{1}{2}\left[u^2\frac{\partial^2}{\partial x^2}p(x,y|t)+v^2\frac{\partial^2}{\partial y^2}p(x,y|t)+2uv\frac{\partial}{\partial x}\frac{\partial}{\partial y}p(x,y|t)\right] \right\}  \ .
\label{eqB9}
\end{eqnarray}
Employing the Eqs.~\eqref{eqB1}--\eqref{eqB7}, the last equation~\eqref{eqB9} becomes
\begin{eqnarray}	   \nonumber
\frac{\partial p(x,y|t)}{\partial t} &=& \frac{\partial^2 p(x,y|t)}{\partial x^2} \frac{1}{2\tau}\int_{-\infty}^{+\infty}\int_{-\infty}^{+\infty}u^2 q(u,v)dudv 	\\ [0.7em]  \nonumber
&+&\frac{\partial^2 p(x,y|t)}{\partial y^2} \frac{1}{2\tau}\int_{-\infty}^{+\infty}\int_{-\infty}^{+\infty}v^2 q(u,v)dudv 	\\ [0.7em]  \nonumber
&+&\frac{\partial}{\partial x}\frac{\partial}{\partial y}p(x,y|t) \frac{1}{\tau}\int_{-\infty}^{+\infty}\int_{-\infty}^{+\infty}uvq(u,v)dudv	\\  [0.7em]
&=&\frac{\langle u^2\rangle}{2\tau}\frac{\partial^2 p(x,y|t)}{\partial x^2} +\frac{\langle v^2\rangle}{2\tau}\frac{\partial^2 p(x,y|t)}{\partial y^2} +\frac{\langle uv\rangle}{\tau}\frac{\partial}{\partial x}\frac{\partial}{\partial y}p(x,y|t) \ .
\label{eqB10}
\end{eqnarray}
Here, the angular brackets indicate averages over the distribution of $u$ and $v$. The last term in Eq.~\eqref{eqB10} is non--zero if the random increments are not independent. In general, to find the particle at the time $t$ at the position $\vec{r}$, the diffusion equation can be written as~\cite{Gupta2010} 
\begin{equation}
\frac{\partial p(\vec{r}|t)}{\partial t}=\nabla\cdot\left(\hat{D}\nabla p(\vec{r}|t)\right) \ .
\label{eqB11}
\end{equation}
In homogeneous and anisotropic media, the diffusion tensor $\hat{D}$ is symmetric and depends on the direction. For a flat two--dimensional space, it is given by
\begin{equation}
\hat{D}=\left[\begin{array}{cc}\hat{D}_{xx} & \hat{D}_{xy} \\\hat{D}_{yx} & \hat{D}_{yy}\end{array}\right]  \ ,
\label{eqB12}
\end{equation}
where $\hat{D}_{xy}=\hat{D}_{yx}$. Thus, the two--dimensional diffusion equation~\eqref{eqB10} turns into
\begin{eqnarray}	   \nonumber
\frac{\partial p(x,y|t)}{\partial t} &=&\left[\begin{array}{cc}\frac{\partial}{\partial x} & \frac{\partial}{\partial y} \end{array}\right]\left[\begin{array}{cc}\hat{D}_{xx} & \hat{D}_{xy} \\\hat{D}_{yx} & \hat{D}_{yy}\end{array}\right] \left[\begin{array}{c}\frac{\partial}{\partial x}p(x,y|t) \\\frac{\partial}{\partial y}p(x,y|t)\end{array}\right]	\\  [0.7em]
&=&\hat{D}_{xx}\frac{\partial^2 p(x,y|t)}{\partial x^2}+\hat{D}_{yy}\frac{\partial^2 p(x,y|t)}{\partial y^2}+2\hat{D}_{xy}\frac{\partial}{\partial x}\frac{\partial}{\partial y}p(x,y|t)  \ .
\label{eqB13}
\end{eqnarray}
Equations~\eqref{eqB10} and \eqref{eqB13} coincide and allow for the identification,
\begin{equation}
\langle u^2\rangle=2\hat{D}_{xx}\tau \ ,
\quad 
\langle v^2\rangle=2\hat{D}_{yy}\tau \ ,
\quad \textrm{and}\quad
\langle uv\rangle=2\hat{D}_{xy}\tau \ .
\label{eqB14}
\end{equation}
For the Brownian motion, the diffusion coefficients $\hat{D}_{xx}$,  $\hat{D}_{yy}$ and $\hat{D}_{xy}$ are constant.

\end{document}